\documentclass[useAMS,usenatbib]{mn2e}
\usepackage{longtable}
\UseRawInputEncoding
\usepackage{graphicx}
\usepackage{subcaption}
\usepackage{lineno}
\usepackage{color}
\usepackage{url}
\usepackage{multirow}
\usepackage{wasysym}
\usepackage{pifont}

\bibpunct{(}{)}{;}{a}{}{,}

\def\Reflex{\textsc{RefleX}}
\def\feka{$\rm {Fe\,K\alpha}$}
\def\nh{$N_{\rm{H}}$}
\def\h2{$H_{\rm{2}}$}
\def\rg{$r_{\rm{g}}$}

\title[The \feka{} Compton shoulder in AGN]{The Fe\,K$\alpha$ Compton shoulder in accreting Supermassive Black Holes}
\author[G. Dimopoulos et al.]{G. Dimopoulos$^{1}$\thanks{E-mail: georgios.dimopoulos@mail.udp.cl}, C. Ricci$^{1,2}$, S. Paltani$^{3}$ \\
\\
$^{1}$Instituto de Estudios Astrof\'{\i}sicos, Facultad de Ingeniería y Ciencias, Universidad Diego Portales,\\ Avenida Ejercito Libertador 441, Santiago, Chile \\
$^{2}$Kavli Institute for Astronomy and Astrophysics, Peking University, Beijing 100871, China \\
$^{3}$Department of Astronomy, University of Geneva, Versoix, Switzerland
 }
\begin{document}
\date{Accepted 2024 July 18. Received 2024 July 8; in original form 2023 June 27}

\pagerange{\pageref{firstpage}--\pageref{lastpage}} \pubyear{2024}

\maketitle

\label{firstpage}

\begin{abstract}
Reprocessed X-ray radiation from active galactic nuclei (AGN) carries important information about the properties of the circumnuclear material around the black hole. The X-ray photons travel from the very center of the system and interact with that material often producing strong emission lines.
The \feka{} Compton shoulder is formed by fluorescent \feka{} photons that perform Compton scatterings with the intercepting material and lose energy to form the distinct shoulder shape.
In this work we use the ray-tracing code \Reflex{} to explore how the physical properties of the medium, as well as its geometry, affect the shape of the Compton shoulder.
We start by running simulations using a simple toroidal reflector, to test the effect of the metal composition, metallicity, column density, dust presence and velocity on the \feka{} line and its Compton shoulder.
We confirm that the shape of the Compton shoulder is sensitive to the optical depth of the intercepting medium, which can be regulated by either changing the metal composition or the line of sight column density of the circumnuclear material.
Next, we create a series of models, which feature different geometrical configurations of dust and gas, and explore how the Compton shoulder is affected by such configurations finding that components that can regulate the line-of-sight column density affect the \feka{} and its CS.
Finally, we test whether observatories such as the recently launched {\it XRISM} and future {\it Athena} will make the Compton shoulder a useful spectral feature of nearby AGN, by applying specific models on simulated spectra of the Circinus galaxy.
The CS has the potential to be used to help constrain properties of the circumnuclear material yet with some limitations.

\end{abstract}	
               

   
\section{Introduction\label{intro}}
Accreting supermassive black holes (SMBH) are known to produce copious amounts of radiation across the whole electromagnetic spectrum (e.g., \citealt{Lynden-Bell:1969}; \citealt{Rees:1984}; \citealt{Brandt:2015}).
X-ray emission is ubiquitously observed in these active galactic nuclei (AGN), the origin of which is suggested to be a hot electron cloud, known as the corona (e.g., \citealt{Matt+:1991,Haardt&mar:1993,Miniutti&Fabian:2004}). 
The size and location of the hot X-ray corona remain uncertain. 
Microlensing variability analysis provides some insights, but it's restricted to a few lensed quasars (e.g., \citealt{Chartas+:2009ApJ}).
Another approach involves studying X-ray spectral-timing in nearby Seyfert galaxies, focusing on X-ray reverberation lags (e.g., \citealt{Fabian:2009,Cackett:2014,DeMarco&Ponti:2019}).
These methods generally support a compact X-ray source near the black hole.
In some Seyfert galaxies, modeling the optical-to-X-ray continuum yields similar coronal size estimates (e.g., \citealt{Done+:2013,Porquet+:2019}).
More recently, using X-ray polarimetry from the Imaging X-ray Polarimetry Explorer (IXPE) \citep{IXPE}, studies have shown that the geometry of the corona can vary between sources \citep{Gianolli+:2023,Ingram+:2023,Tagliacozzo+:2023}.

Optical/UV photons produced in the accretion disk are up-scattered by the corona into the X-ray band. This X-ray radiation can then be reprocessed by the circumnuclear environment of the SMBH (e.g., \citealt{George&Fabian:1991}), for example the accretion disk or the dusty torus \citep{Antonucci:1993}.
This reprocessed radiation is the result of various interactions between the seed X-ray photons and the intercepting medium before collected by X-ray observatories.
The footprint of these interactions on the spectrum is what carries information about the aforementioned medium and hence, its study can provide important information about the physical conditions and geometry of the close environments of SMBHs (e.g., \citealt{Antonucci:1993,Urry&Padovani:1995,RamosAlmeida&Ricci:2017Nat}).

Fluorescent emission lines (e.g., $\rm Fe\,K\alpha\ at\ 6.4\, keV$) are among the most characteristic features of reprocessed X-ray radiation observed in AGN. The probability of the emission of a fluorescent line photon is controlled by the fluorescent yield ($Y$).
The \feka{} emission line is possibly the most common feature in the X-ray spectra of AGN due to the relatively high abundance of Fe, and the fact that $Y$ increases rapidly with the atomic number ($Y \propto Z^4$).
Fluorescent photons after their emission can interact further with the intercepting medium. Free electrons, bound to neutral or partially ionized electrons are targets of the photons.
When Compton scattering occurs between the photons and these "cold" electrons a low energy tail is formed in energies just bellow the fluorescence peak, a phenomenon that should not be confused with the up-scattering that takes place in the hot corona where low energy photons from the disk collide with highly energetic electrons and gain energy through that process.
This low energy tail is a feature known as the Compton shoulder (CS hereafter).
Due to the limitations of current X-ray facilities, this features has been detected only in a handful of sources (e.g., \citealt{Bianchi:2002,Shu+Yaqoob+:2011TypeII}).
A prominent \feka{} CS has been detected in the high mass X-ray binary (HMXB) GX\,301-2 \citep{Sato:+1986,Koh_1997,Nabizadeh+:2019}. Using {\it Chandra}-HETG, 
\citet{Watanabe+:2003} fit the CS to estimate the column density, electron temperature and metal abundance of the spherical cloud material that surrounds this X-ray pulsar, which has shown the potential of the CS as a spectral feature that can be useful to constrain the material around accreting objects. \citet{Torrejon:2010} studied 31 X-ray binaries (XRBs) using spectra taken by {\it Chandra}, focusing on the \feka{} region, and revealed the presence of a \feka{} CS in the HMXB X1908+075.
The detection of the CS in AGN is rather rare, because of the limited spectral resolution and sensitivity of the current observatories.
\citet{Hikitani:2018} worked with {\it Chandra}-HETG data of the {\it Circinus} galaxy. They were able to resolve the CS of the \feka{} and use it to extract some information about the properties of the obscuring material.

Simulations can be a powerful tool to establish the link between the \feka{} CS and the properties of the circumnuclear material of the SMBH.
A variety of works has been done considering different geometries and physical properties of the circumnuclear environment. \citet{Matt:2002} tested the dependence of the CS on the column density and inclination angle, assuming that the reprocessing material is a sphere or a slab.
\citet{Murphy&Yaqoob:2009} performed Monte Carlo simulations considering that the X-ray source is surrounded by a torus.
They also analysed the \feka{} line and the ratio between the fluorescent photons and the scattered ones, which form the CS, finding that for low column densities ($ N\rm_H < 10^{23}\,cm^{-2}$) the ratio of the scattered photons over the line photons is very low ($\sim 2\%$).
However, as the \nh{} increases that fraction increases as well, up to $ N\rm_H \sim 10^{24}\, cm^{-2}$, and then it drops due to the high absorption.
\citet{Yaqoob&Murphy:2010} performed similar simulations, and showed that the shape of the CS is sensitive to changes on the column density of the intercepting material, especially in high observing angle.
For \nh{} ($> 5\times10^{24}\, \rm{cm^{-2}}$) the CS shows a peak at $6.24\,\rm{keV}$, which is the maximum energy shift for 1x scattered photons due to electron recoil for photons back-scattered, $\Delta E_{\rm{max}} = 2E_{\rm{\circ}} ^2 / (m_{\rm{e}} c^2 + 2 E_{\rm{\circ}})$.
They explain that this is related to the incapability of photons to penetrate deep into the torus at high optical depths.
It was also suggested that flatter spectra, which can provide higher amount of energetic photons, can travel deeper into the medium, so that the spectral index of the incident photon field can also affect the shape of the CS.
\citet{Yaqoob&Murphy:2010} also examined how the velocity broadening of the \feka{} line can blend in with the CS, and illustrated that for velocities $\rm{FWHM > 2000\, km\,s^{-1}}$, it becomes hard to distinguish the CS feature, with its only discernible indication being a subtle asymmetry within the blended line profile. Consequently, CCD X-ray detectors lack the capability to identify the Compton shoulder since they feature spectral resolution FWHM of the order of $\sim 7000\,\rm{km\,s^{-1}}$ or more.

\citet{Furui:2016} used the \texttt{MONACO} simulation platform \citep{Odaka+:2011} and studied the dependence of the CS on different torus parameters.
They found a strong relation between the CS intensity and the inclination angle, column density and metal abundance of the torus.
They showed that the ratio between the iron line and its CS depends on the various factors tested like the physical properties or the configuration of the torus and whether it is a smooth or clumpy distribution of material.
\citet{Odaka:2016} investigated how the shape of the CS changes, by using Monte Carlo simulations for spherical and slab geometries of the circumnuclear material.
In the case of the spherical medium, they performed simulations testing the column density of the cloud, its metallicity and the spectral slope of the intrinsic continuum.
The shape and flux of the CS vary with the physical properties of the sphere (\nh{}, metallicity $Z$), whereas the photon index ($\rm{\Gamma}$) has no influence on it which can be, however, a result of the moderate column density tested.
In the case of the slab reflector they explored the angular dependence of the shape of the CS and showed that the shape of the CS at different angles is characterized by the dominant scattering angle in each case.

The Compton shoulder has not been established yet as an observational feature to probe the properties of the circumnuclear material, due to the lack of high-resolution X-ray observatories.
The new generation of X-ray spectrographs, such as the one onboard the recently-launched {\it XRISM} observatory \citep{xrism:2020}, will have the necessary spectral resolution and sensitivity to resolve the CS of AGN.
In this work, we perform ray-tracing simulations to explore how the physical and geometrical properties of the reprocessing material affect the shape and flux of the \feka{} CS. To do this we use the ray-tracing platform \Reflex{} \citep{Paltani&Ricci:2017}. \Reflex{} gives us the freedom to set different physical conditions as well as complex geometries.
Many different properties have been tested such as the composition of metals of the medium(\citealt{Anders:1989,Wilms:2000,Lodders:2003}),
the abundance of it (metallicity, $Z$), the column density (\nh{}) and different dust abundances \citep{Ricci&Paltani:2023}.
Moreover, the effect of the velocity broadening has been investigated as well as properties of the X-ray source.
Besides the physical properties, we use different geometrical objects to form various configurations of the circumnuclear material, by including components such as the accretion disk, the dusty torus, the broad line region (BLR) and a polar hollow cone. 
In \S\ref{reflex} we present the \Reflex{} platform and the setup of our simulations.
Then, in \S\ref{building_blocks} we describe the geometrical objects
we use within our simulations. In \S\ref{results} we present the results of our simulations organized as following. First, we talk about the physical properties of a toroidal structure and how it affects the CS and then we present the different models of the circumnuclear medium based on the different geometries mentioned above.
In \S\ref{sec:simulations} we test the models we have developed on simulated spectra of the next generation of X-ray observatories
\textit{XRISM} and {\it Athena}.
Finally, the conclusions of our study are presented in \S\ref{summary}.

\section{Simulations}\label{sim_setup}

\subsection{RefleX}\label{reflex}
To build a physically driven model of AGN and test the Compton shoulder, we simulate X-ray spectra using the ray-tracing code \Reflex{} \citep{Paltani&Ricci:2017,Ricci&Paltani:2023}.
\Reflex{} simulates the propagation of X-ray photons from $\rm0.1\, keV$ to about $\rm1\, MeV$, and includes the majority of the continuum processes in cold material that occur at these energies.
Moreover, it introduces a modular structure, that gives the option to the user to test models by adding multiple building blocks.
These building blocks can vary in their geometry and physical properties.
Using Monte Carlo simulations, \Reflex{} tracks each photon, recording all the interactions that occur as it travels through the material.
The code provides images and spectra for the geometrical setup chosen by the user.
In our study, we considered a disk, sphere, torus and hollow cone as building blocks.

\subsection{Building blocks}\label{building_blocks}

\subsubsection{The X-ray source}\label{sec:xsource}
The first component of our simulations is the X-ray source.
It is widely accepted that X-ray photons originate in a hot plasma cloud, called corona.
The geometry and the position of the corona is still under debate, nevertheless we adopt the lamp-post model which places a spherical corona a few gravitational radii on top of the SMBH (e.g., \citealt{Fabian:2009,emmanoulopoulos+:2014,uttley+:2014}).
The location of the corona is fixed at $ 10\,r_{\rm{g}}$ ($ r_{\rm{g}} = 2GM_{\rm{BH}}/c^2$ is the gravitational radius) and its radius at $6\,r_{\rm{g}}$ (e.g., \citealt{Chartas:2009,Fabian:2009,DeMarco:2013}).
In order to calculate the distance and the size of the corona,  the black hole mass of the Circinus galaxy was used, $\log\left(M/M_{\odot}\right) = 6.23$ \citep{Koss+:2017}.
The shape of the X-ray source can however differ from the typical spherical corona, due to different accretion mechanisms or even the effect of general relativity on the electron population (e.g., \citealt{Ptak+:1998ApJ}).
In Appendix \ref{App:xray_source_geom} we test different geometries of the X-ray source to explore their effect on the shape of the CS.
We have tested besides the corona described above, a point source as well as an ADAF spherical cloud. None of the tested geometries change the shape of the CS.
Yet, it is important to mention that GR effects near the source has not been included in the current work that can affect the overall shape of the spectrum (e.g. \citealt{GarciaGR:2014}).

The properties of the input X-ray photons are fixed throughout our study.
We generate photons in the energy range $\rm{5-50\,keV}$ to cover the part of the X-ray spectrum known to be dominated from reflection effects, including the \feka{} and its Compton shoulder.
The energy binning is, also, fixed at $\rm{3\,eV}$, which provides a spectral resolution sufficient to study the CS in detail.
The photon distribution selected is a power law with photon index $\Gamma=1.8$ and energy cutoff $E_{\rm{C}}=200\,\rm{keV}$, in agreement with recent studies of bright nearby AGN (e.g., \citealp{Ricci+:2017Catalog,Ricci+:2018}).
In previous studies it was tested whether the photon index can affect the shape of the CS (e.g., \citealt{Yaqoob&Murphy:2010,Odaka+:2011}).
To explore whether various source properties can affect the shape of the \feka{} and its corresponding CS, in Appendix\,\ref{App:XsourceGamma} we compare flat and steep spectra for two different toroidal configurations.
In our simulations the results of the different spectral indices on the CS is
mostly found on the flux of it and not at the shape of it (see Fig.\,\ref{fig:torus_gamma} which suggests that the CS as spectral feature does not show high potential in constraining the spectral properties of a given source. 

To convert the photon counts to observable spectra we fix the AGN luminosity at $L_{2-10} = 3\times 10^{42}\,\rm{erg\,s^{-1}}$, which corresponds to the intrinsic $\rm 2-10\,keV$ luminosity of the Circinus galaxy \citep{Arevalo+:2014}, and the redshift to $z = 0.001$ \citep{Koribalski+:2004}.
In Table\,\ref{tab:source_details} the details of the X-ray source are reported.

\begin{table}
\centering
\caption{The properties of the X-ray source and the photon distribution of the input radiation field. The source is placed above the SMBH.}
\begin{tabular}{c c}
\hline
\hline
\multicolumn{2}{c}{\multirow{2}{*}{\textbf{\large{Black Hole}}}} \\
\\
Mass & $\rm 10^{6.23}\,M_{\odot}$ \\
redshift & 0.001 \\
\hline
\multicolumn{2}{c}{\multirow{2}{*}{\textbf{\large{X-ray source}}}} \\
\\
Geometry & Sphere \\
Distance & $10r_{\rm{g}}=1.62\times 10^{-6}\,{\rm pc}$ \\
Radius & $6r_{\rm{g}}=9.75\times 10^{-7}\,{\rm pc}$ \\
$\Gamma$ & 1.8 \\
$E_{\rm c}$ & $200\,{\rm keV}$\\
Band & 5 - 10 keV \\
Binning & 3 eV \\
\hline
\end{tabular}

    \label{tab:source_details}
\end{table}

\subsubsection{The accretion disk}\label{sec:AD}

In order to explore how X-rays interact with the environment of the SMBH, we need to define different distributions of material around the black hole.
The SMBH is surrounded by an accretion disk (AD) that may extend to several hundreds \rg{} (e.g., \citealt{Jha+:2022}).
We created a disk component that starts at $r_{\rm{inner}}=10r_{\rm{g}}$ and terminates where the broad line region (BLR) begins at $r_{\rm{BLR}}=0.0037\, {\rm pc}$ (see\,\S\ref{sect:BLR}).
Moreover, the AD is considered to be opaque, and therefore we set the volumetric density at $\rm 10^{12}\,cm^{-3}$ (e.g., \citealp{Garcia+:2013}). Hydrogen and helium were set to be fully ionized, similarly to the prescriptions used for the \textsc{pexrav} model \citep{Magdziarz&Zdziarski:1995}. 
The metallicity is set to one and the default abundance used is \texttt{lodd} \citep{Lodders:2003}.
In our simulation general relativity (GR) effects from the area close to the SMBH are not included and the study of GR on X-ray reflection is outside the scope of this work (see details in \citealt{GarciaGR:2014}.
Nevertheless, the accretion disk and its relative position in the system can work as a giant mirror of cold material which is interesting to explore whether it contributes to the formation of the CS and therefore we decided to implement it.
In \S\ref{sec:velbroad} we show that when any kind of broadening effects are applied to the spectrum the CS becomes less useful as a tool for the study of the circumnuclear medium.

In\,Fig.\,\ref{fig:cartoon} a cartoon of all the components is presented, including 
the accretion disk (AD), the properties of which are presented in Table\,\ref{tab:component_details}.

\subsubsection{The broad line region}\label{sect:BLR}
The broad line region (BLR), i.e. the region where broad optical/UV lines are produced, is located after the accretion disk and extends up to the sublimation radius \citep{Suganuma+:2006,Davies+:2015}, where dust is formed. 
We model it using a flared disk geometry (e.g., \citealt{GravityCollab:2018Natur}, \citeyear{GravityColab:2020}). 
To build the flared disk, we used 20 annuli blocks with decreasing inner radius along the vertical axis (see\,Fig.\,\ref{fig:cartoon}).
In Appendix \ref{App:BLRblocks} we test different numbers of blocks to recreate the flared disk geometry for the BLR.

The abundance of the BLR is set to \texttt{lodd}, the fraction of molecular hydrogen is set to $20\%$ ($H_{\rm{2}}=0.2$), the volumetric density is fixed for all the annuli at $\rm 10^{8}\,cm^{-3}$ which corresponds to equatorial column density $\sim 10^{25}\,cm^{-2}$ (e.g., \citealp{Ferland+:1992,Goad+:2012}) and the gas is set to be neutral.
The size of the BLR has been calculated considering that its outer radius corresponds to the sublimation radius.
First, the sublimation radius was calculated using the following formula \citep{Mor:2009}:
\begin{equation}
    R_{\rm{sub}}=1.3L^{0.5}_{46} \times \left( \frac{1500\rm{K}}{T_{\rm{sub}}} \right)^{2.6}\,\rm{pc}   
\end{equation}

where $R_{\rm{sub}},\ T_{\rm{sub}}$ are the sublimation radius and temperature, respectively, while 
$L_{46} = L_{\rm{bol}}/10^{46}\rm\,erg\,s^{-1}$. 
We adopted the $\rm{2-10\,keV}$ luminosity of the Circinus galaxy ($L_{2-10} = 3\times 10^{42}\,\rm{erg\,s^{-1}}$, \citealp{Arevalo+:2014}), and
converted the X-ray luminosity to bolometric using as bolometric correction $\kappa=20$ \citep{Vasudevan:2009}.
The inner radius was set to \citep{Kaspi:2005}:
\begin{equation}
    \frac{R_{\rm{BLR}}}{10\,\rm{lt\,days}} = 0.86\times \left( \frac{L_{2-10}}{10^{43}\,\rm{erg\, s^{-1}}} \right)^{0.544}\,\rm{pc}   
\end{equation}

In Table\,\ref{tab:component_details} the properties of the BLR are summarized.

\subsubsection{The torus}\label{sec:torus}

The "torus" is an anisotropic and possibly axisymmetric structure that is found outside the sublimation radius, which was originally proposed in early AGN unification models (e.g., \citealt{Antonucci:1993,Urry&Padovani:1995}).
This structure is responsible for obscuring the central region of AGN (e.g., \citealt{Awaki+:1991,Guainazzi+:2005,Guainazzi+:2016}) and reflecting an important amount of radiation (e.g., \citealt{Matt+:1991,Panagiotou&Walter:2019}).
In\,Fig.\,\ref{fig:cartoon} we illustrate the toroidal structure and its position in the system. We use the \texttt{lodd} abundance and set the molecular hydrogen fraction to $H_2 =0.3$ and the volumetric density to $\sim 2.2\times 10^{6}\,\rm{cm^{-3}}$, which corresponds to column density $N_{\rm{H}} = 10^{24.5}\,\rm{cm^{-2}}$.
The material is assumed to be neutral.
The inner radius is fixed to the sublimation radius, since the torus is expected to be dusty.
Hence, the iron depletion factor is set to 70\% (DUST=0.7 in \Reflex{} notation).
Moreover, the covering factor is $CF = 0.7$, consistent with the results of recent surveys of nearby AGN (e.g., \citealp{Ricci+:2015,Ricci+:2017Natur,Ricci+:2022}).
In Table\,\ref{tab:component_details}, we list the properties of the torus.
 
\subsubsection{The cone}

\begin{figure}
    \centering
    \includegraphics[width=\columnwidth]{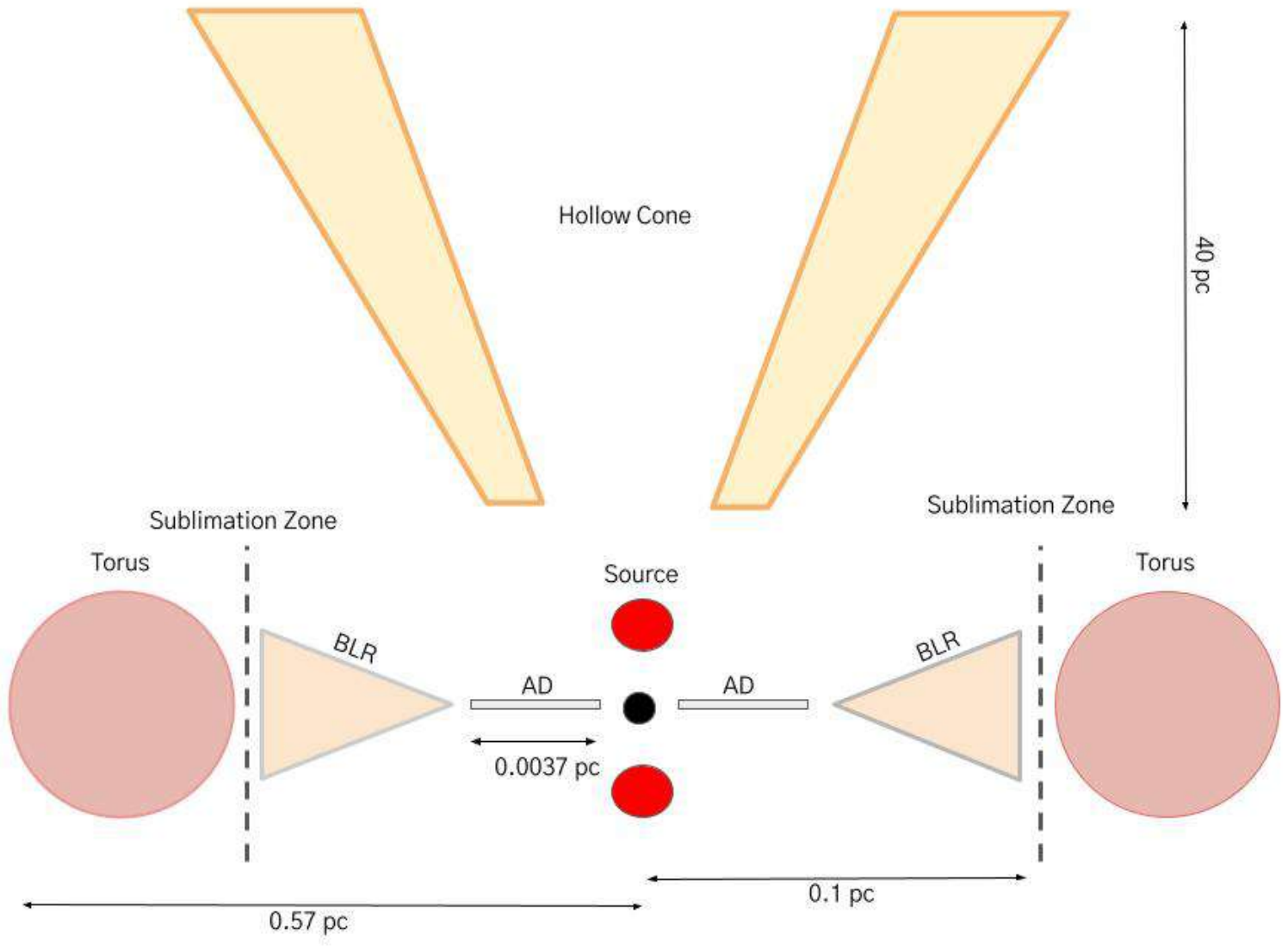}
    \vspace{-3mm}
    \caption{Different geometrical components that were considered in our simulations and modelling. Throughout this study various geometrical objects are used. The complete model, composed of all the geometrical objects is the model M6 as described in \S\ref{geometries}. The objects are not scaled.}
    \label{fig:cartoon}
\end{figure}

Infrared interferometry and single-dish observations of nearby AGN have discovered a warm ($\sim 300 - 400\,\rm{K}$) polar dust structure, and a significant fraction of the AGN mid-infrared (MIR) emission could actually be due to this component \citep{Hoenig+:2014,Asmus+:2014,Asmus:2019}.
In order to implement this polar medium in our simulations, we use a hollow cone that extends up to a few parsecs from the SMBH, a geometry that has been reported for the Circinus galaxy (e.g., \citealt{Stalevski+:2017,Stalevski+:2019}).
In\,Fig.\,\ref{fig:cartoon} the polar material is illustrated as a yellow cone.
The cone starts at $\rm 0.18\,pc$, which is the distance of the sublimation radius, equal to the one used for the torus, and extends up to $\rm 40\,pc$.
Its half opening angle is $40^{\circ}$ with outer angle $+10^{\circ}$, following \citet{Stalevski+:2017}. The volumetric density of the hollow cone is $\sim 62.1\rm\,cm^{-3}$, which corresponds to $N_{\rm{H}}\sim 10^{22}\rm\,cm^{-2}$. The material in the cone is neutral and has a metallicity $Z=1$.
Moreover, we set the molecular hydrogen $H_{\rm{2}} =0.4$ (e.g. \citealp{Wada+:2009}).
Like the torus the polar cone is expected to be dusty and therefore the dust was set to 0.7 too.
In Table\,\ref{tab:component_details} the physical properties of the hollow cone are presented in detail.

\begin{table}
\centering
\caption{Properties of the different components that are used throughout our simulations.}
\begin{tabular}{c c}
\hline
\hline
\multicolumn{2}{c}{\multirow{2}{*}{\textbf{\large{Accretion Disk}}}} \\
\\
Geometry & Disk \\
Inner Radius & $\rm 1.625\times 10^{-6}\, pc$ \\
Outer Radius & $\rm 0.0037\, pc$ \\
Density & $\rm 10^{12}\, cm^{-3} $ \\
State & H, He ionized \\
$H_2$ & 0 \\
$Z$ & 1 \\
\hline
\multicolumn{2}{c}{\multirow{2}{*}{\textbf{\large{Broad Line Region}}}} \\
\\
Geometry & Flared Disk \\
Inner Radius & $\rm 0.0037\, pc$ \\
Outer Radius & $\rm 0.1\, pc$ \\
Density & $\rm10^{8}\, cm^{-3}$ \\
$CF$ & 0.4 \\
State & neutral \\
$H_2$ & 0.2 \\
$Z$ & 1 \\
\hline
\multicolumn{2}{c}{\multirow{2}{*}{\textbf{\large{Torus}}}} \\
\\
Geometry & Torus \\
Inner Radius & $\rm 0.1\, pc$ \\
Outer Radius & $\rm 0.57\, pc$ \\
$CF$ & 0.7 \\
Density & $\rm 2.18\times 10^{6}\, cm^{-3}$ \\
\small{(\textbf{Column density}} & \small{$N_{\rm H}= 10^{24.5}\, \rm cm^{-2}$)} \\
State & neutral \\
Dust & 0.7 \\
$H_2$ & 0.3 \\
$Z$ & 1 \\
\hline
\multicolumn{2}{c}{\multirow{2}{*}{\textbf{\large{Hollow Cone}}}} \\
\\
Geometry & Cone \\
Bottom & $\rm 0.1\, pc$ \\
Top & $\rm 40\, pc$ \\
Angle (width) & $\rm 40^{\circ}\ (10^{\circ})$ \\
Density & $\rm 62.1\, cm^{-3}$ \\
State & neutral \\
Dust & 0.7 \\
$H_2$ & 0.4 \\
$Z$ & 1 \\
\hline
\end{tabular}

\label{tab:component_details}
\end{table}

\section{Results}\label{results}

In this section we present the results of our simulations, which can be divided into two categories.
The first set of simulations is focused on exploring different physical parameters of the circumnuclear material and their effect on the CS. For this, we used a simple model to avoid possible degeneracies between the physical conditions and the geometry of the system. The X-ray source is spherical, as introduced in \S\ref{sec:xsource} (see, Table\,\ref{tab:source_details}), while the reprocessing medium used is a \textit{torus only} model (see\,Table\,\ref{tab:component_details}).
The properties we tested are: i) metal abundance/composition, ii) metallicity, iii) column density, iv) dust depletion, v) velocity broadening.
We decided to refer to metal abundance or composition to the table of metals-elements that correspond to command \texttt{abund}\footnote{https://heasarc.gsfc.nasa.gov/xanadu/xspec/manual/node116.html} in XSPEC.
On the other hand, the metallicity refers to the factor that is multiplied with the densities shown in the aforementioned tables.
Moreover, in Appendix \ref{App:h2fraction} we explored whether the presence of molecular hydrogen can affect the shape of the CS, showing that its effect is negligible. For the second part of our work we fixed the physical properties of the geometrical objects, considering the values reported in Table\,\ref{tab:component_details}. Different geometrical objects were included into our simulations, and their output spectra were compared, in order to explore the diversity in flux and shape of the \feka{} line and its CS.

\Reflex{} includes the option to select photons based on the type and number of interactions they have undergone from the moment they are generated to the moment they are collected.
The scope of this work is to investigate the effect of the circumnuclear material on the \feka{} line and its CS.
Hence, we select photons that originate in fluorescence ($\rm FLUOR > 0$ in \Reflex{} notation). 

The simulated system can be observed in various observing angles via \Reflex{}.
However, throughout our main study we have deliberately limited our analysis in the edge-on scenario ($85^{\circ} - 90^{\circ}$ angle width).
In previous studies (e.g. \citealt{Yaqoob&Murphy:2010}) it has been shown that the observing angle can play a role to the observed CS shape and moreover,
it is not necessarily a known property in various system, and the different components (e.g. disk and torus) are not always aligned.
Nevertheless, in the core analysis of the behavior of the CS we have decided to keep the inclination angle of the system fixed to edge on.
Studies of masers suggest a correspondence between high inclination angles and large column densities (i.e. $N_{\rm H}\geq 10^{23}\,{\rm cm^{-2}}$), which is also necessary for a strong CS (see \S\ref{col_density}), (e.g., \citealt{Masini+_megamasersNUSTAR:2016A&A,Panessa+waterMasers:2020A&A}).
Therefore, we decided to keep the inclination angle with respect to the system fixed, and corresponding to a edge-on view, throughout our analysis.
However, the reader should be aware of the degeneracies that can be introduced by a varying observing angle.
To illustrate it, in Appendix\,\ref{App:Inclination} we present the \feka{} and CS energy region for different inclination angles.
To sum up, unless it is specifically mentioned, the default observing angle bin throughout the main analysis is $85^{\circ} - 90^{\circ}$.

The effect of the aforementioned properties can be quantified in terms of flux changes on the \feka{} emission line ($F_{\rm{K\alpha}}$) and its corresponding CS  ($F_{\rm{CS}}$).
To quantify this, we measure the flux of these two components (6.230 up to $\rm6.380\, keV$ and from 6.385 to $\rm6.42\, keV$ for the CS and \feka{} line, respectively), and compare their ratio ($F_{\textrm{CS}}/F_{\textrm{FeK$\alpha$}}$), to study how different physical properties affect the \feka{} line and its CS (see also \citealt{Odaka:2016,Furui:2016,Hikitani:2018}). 
All the fluxes are calculated with the continuum subtracted, since we collect only reprocessed photons.

Besides flux changes, the CS can show variations in its shape (e.g., asymmetry). 
First, to quantify the asymmetry we divide the CS in two branches left and right.
The flux of the left branch is integrated from 6.230 up to $\rm 6.305\, keV$ and the right one from 6.305 to $\rm6.38\, keV$ and calculate their ratio (asymmetry ratio).
Finally, we check how the CS flattens or dents by changing the parameters of the reflecting material. 
To do that we consider energy of the \feka{} photons at $\rm \sim 6.24\, keV$ as the peak energy of "back-scattering", for the edge-on regime (see also Appendix\,\ref{App:Inclination}, and the lowest flux point in energy range between this "peak" and the emission line ($\rm 6.28-6.38\, keV$) as "bottom".
We calculate the $peak/bottom$ ratio to make an estimation of the shape of the CS. Note that the peak at $\rm \sim 6.24\, keV$ is valid for edge-on observations, which is the default configuration. In other observing angles the dominant peak of the CS can significantly deviate from this energy (see\,Appendix\,\ref{App:Inclination}).

\subsection{Physical properties}\label{sec:physical_prop}

\subsubsection{Metal composition}\label{abnd}

The first property tested is the metal composition of the torus. We ran a series of simulations to examine how different compositions affect the \feka{} emission line and its CS. To do so, we use the option of \Reflex{} that allows us to select between the most common ones used in the literature.
Following \textsc{XSPEC} \citep{XSPEC} notation\footnote{\scriptsize{https://heasarc.gsfc.nasa.gov/xanadu/xspec/manual/node115.html}} we tested: i) \texttt{angr} \citep{Anders:1989}, ii) \texttt{wilm} \citep{Wilms:2000} and iii) \texttt{lodd} \citep{Lodders:2003}.
The \texttt{angr} composition is the one with the highest amount of metals while it has significantly more iron ($4.68\times 10^{-5}$, units in comparison to Hydrogen) than the other two.
The \texttt{lodd} and \texttt{wilm} compositions have similar Fe abundances, $2.69\times 10^{-5}$ and $\ 2.95\times 10^{-5}$ respectively, while \texttt{lodd} is "richer" in almost every other element.
In general, selecting the metals available in the medium is a way to tune the overall absorption of the incident photons.
The total cross-section of the medium changes and for a given column density this results into different optical depth.
As a result, the metal composition is expected to have an influence on the \feka{} and CS.

\begin{figure}
    \centering
    \begin{subfigure}{\columnwidth}\includegraphics[width=\columnwidth]{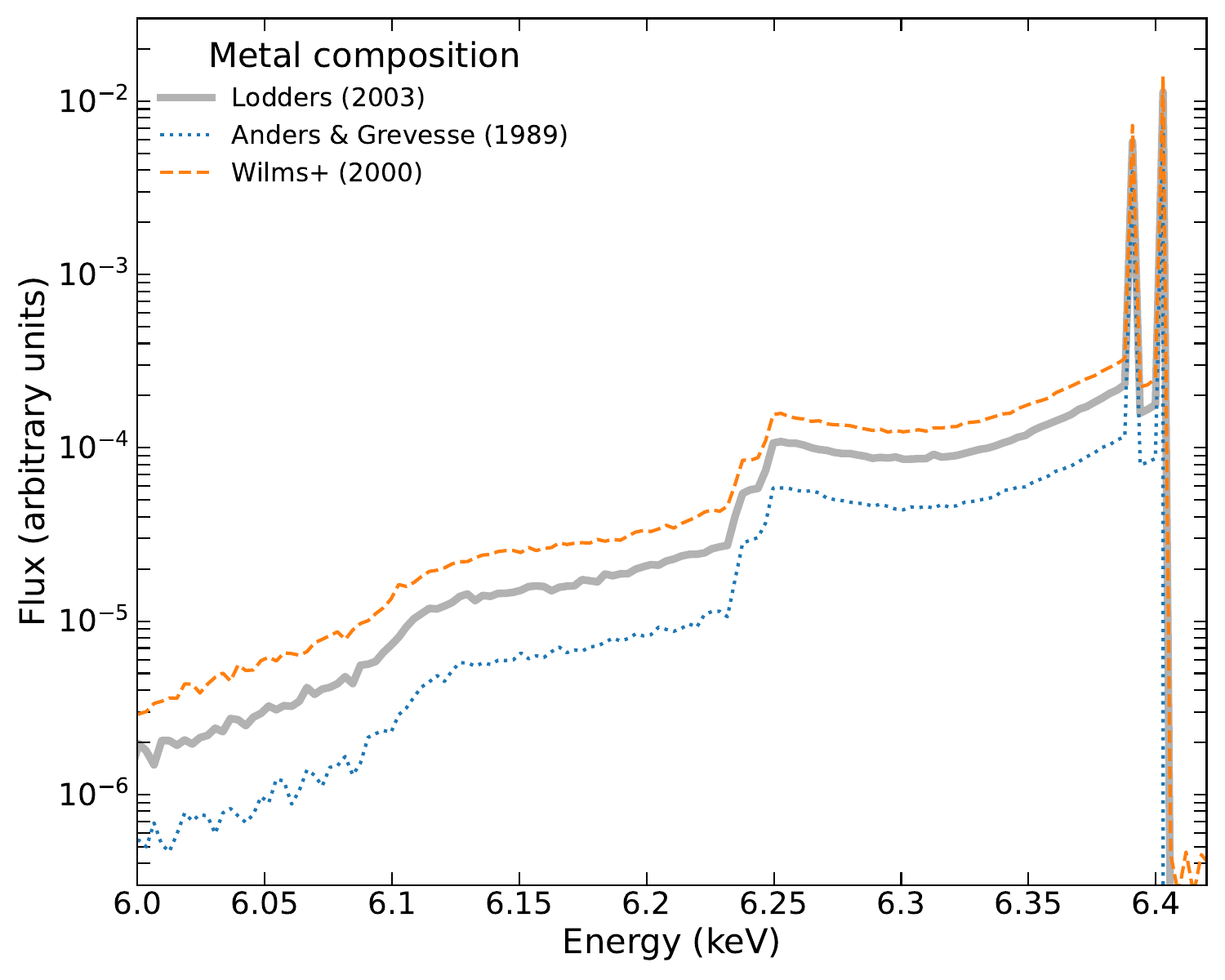}
    \hfill
    \end{subfigure}
    \begin{subfigure}{\columnwidth}\includegraphics[width=\columnwidth]{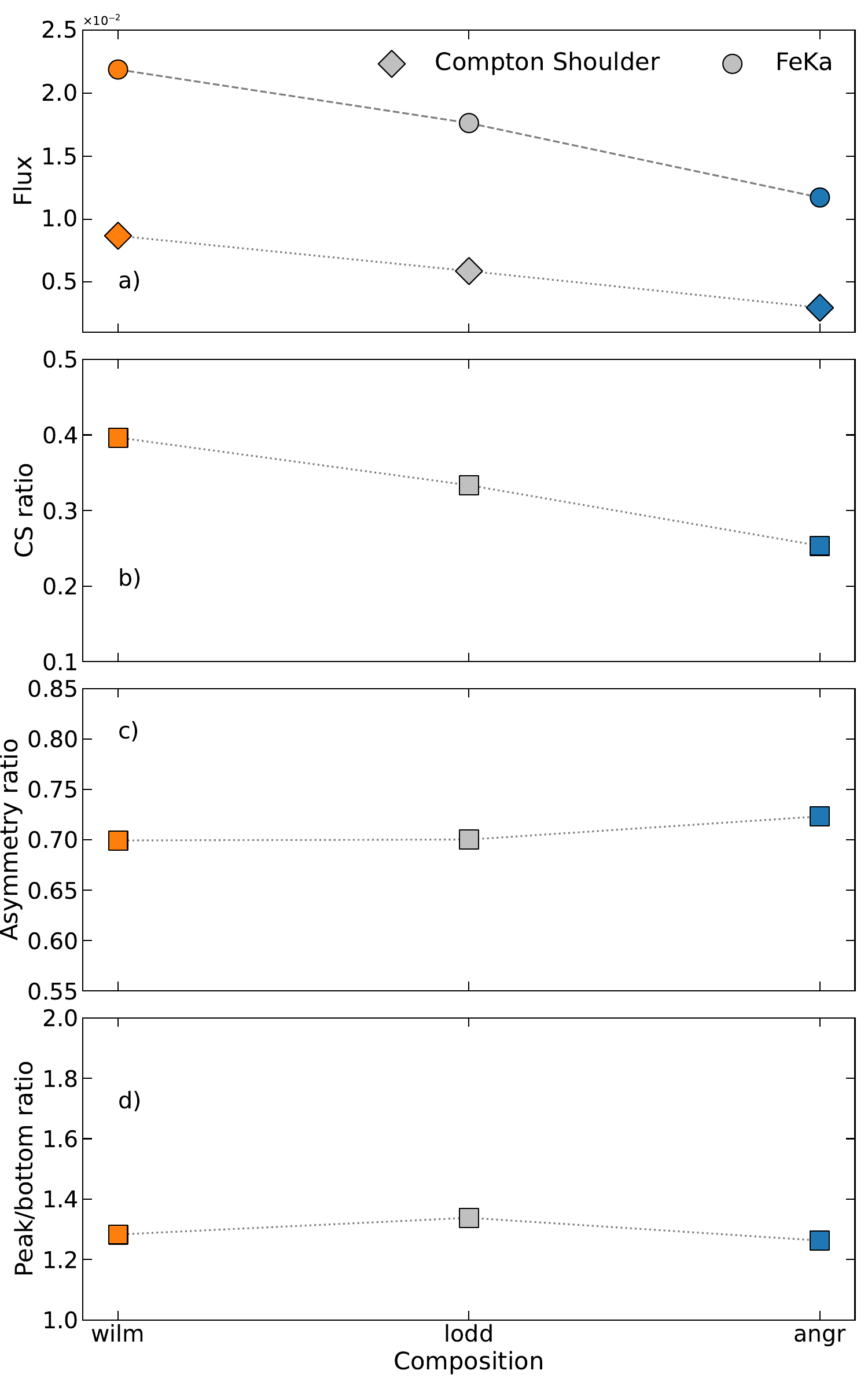}
    \end{subfigure}
    \vspace{-3mm}
    \caption{{\textit{Top}}: The Compton shoulder for different metal abundances. The {\sc xspec} notation is used here. In all examples we consider a simple torus, observed edge on, and a spherical corona as the X-ray source. {\textit{Bottom}}: Top to bottom the four panels show: a) the flux of the CS and \feka{} b) flux ratio CS/\feka{} c) asymmetry ratio and d) peak/bottom ratio.}
    \label{fig:torus_abnd}
\end{figure}

In\,Fig.\,\ref{fig:torus_abnd}-top we present the output of the simulations for the three cases described above. For the \texttt{wilm} abundances the spectrum has the highest flux, followed by \texttt{lodd} and finally by \texttt{angr}. 
The \texttt{wilm} abundances, which has the lowest amount of metals, results in less absorption and therefore, the highest overall flux.
This can be better noticed in panel (a) of Fig.\,\ref{fig:torus_abnd} where the fluxes of both the \feka{} and its CS are presented.
Both features show a negative trend which implies that having more metals, increases the overall absorption.
The \feka{} line decreases in terms of flux despite the presence of more iron as the balance between more \feka{} emission and absorption favors the absorption.
To further test this effect, we have run a series of simulations with the same composition - \texttt{lodd} - but variable amount of iron. We found that for Fe abundances from $10^{-2} \times$ {\it table value} up to $10^{3} \times$ {\it table value} the overall \feka{} and CS flux drops due to high absorption.
In panel (b) of Fig.\,\ref{fig:torus_abnd} we illustrate the ratio of the two fluxes which shows a slight negative trend but only changes for 0.15.
The asymmetry of the three compositions is almost identical which suggests that the different element tables regulate the "amplitude" of the CS but do not affect its shape, which is also supported by what we found for the peak/bottom ratio, which slightly changes as well (see panels (c) and (d) of Fig.\,\ref{fig:torus_abnd}).

In the following we adopt the \texttt{lodd} metal abundance as the default option throughout this study (grey line in\,Fig.\,\ref{fig:torus_abnd}), since it provides sufficient overall flux and produces a significant Compton shoulder feature (see also Table\,\ref{tab:component_details}).

\subsubsection{Metallicity ($Z$)}\label{metal}

\begin{figure}
    \centering
    \begin{subfigure}{\columnwidth}\includegraphics[width=\columnwidth]{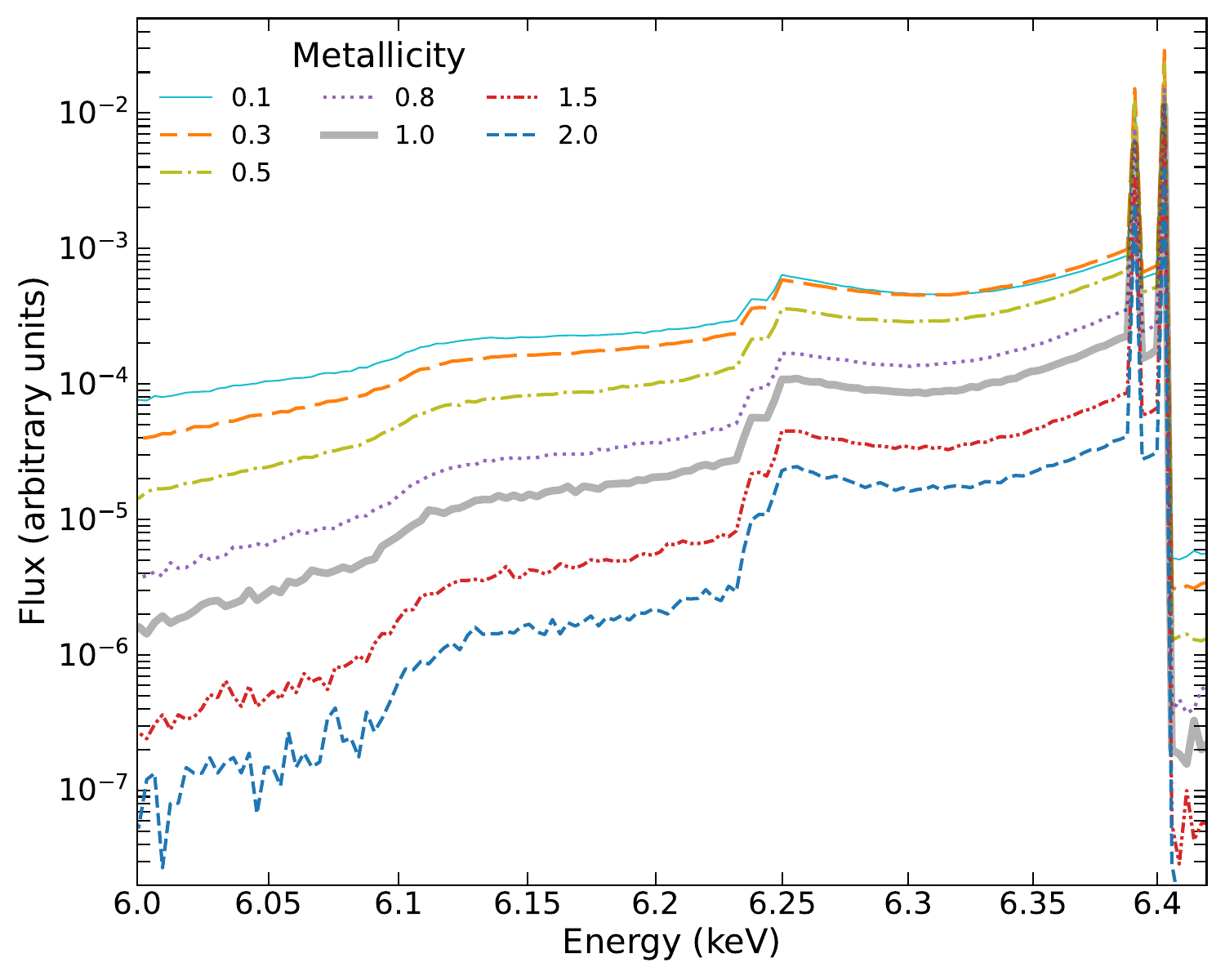}
    \hfill
    \end{subfigure}
    \begin{subfigure}{\columnwidth}\includegraphics[width=\columnwidth]{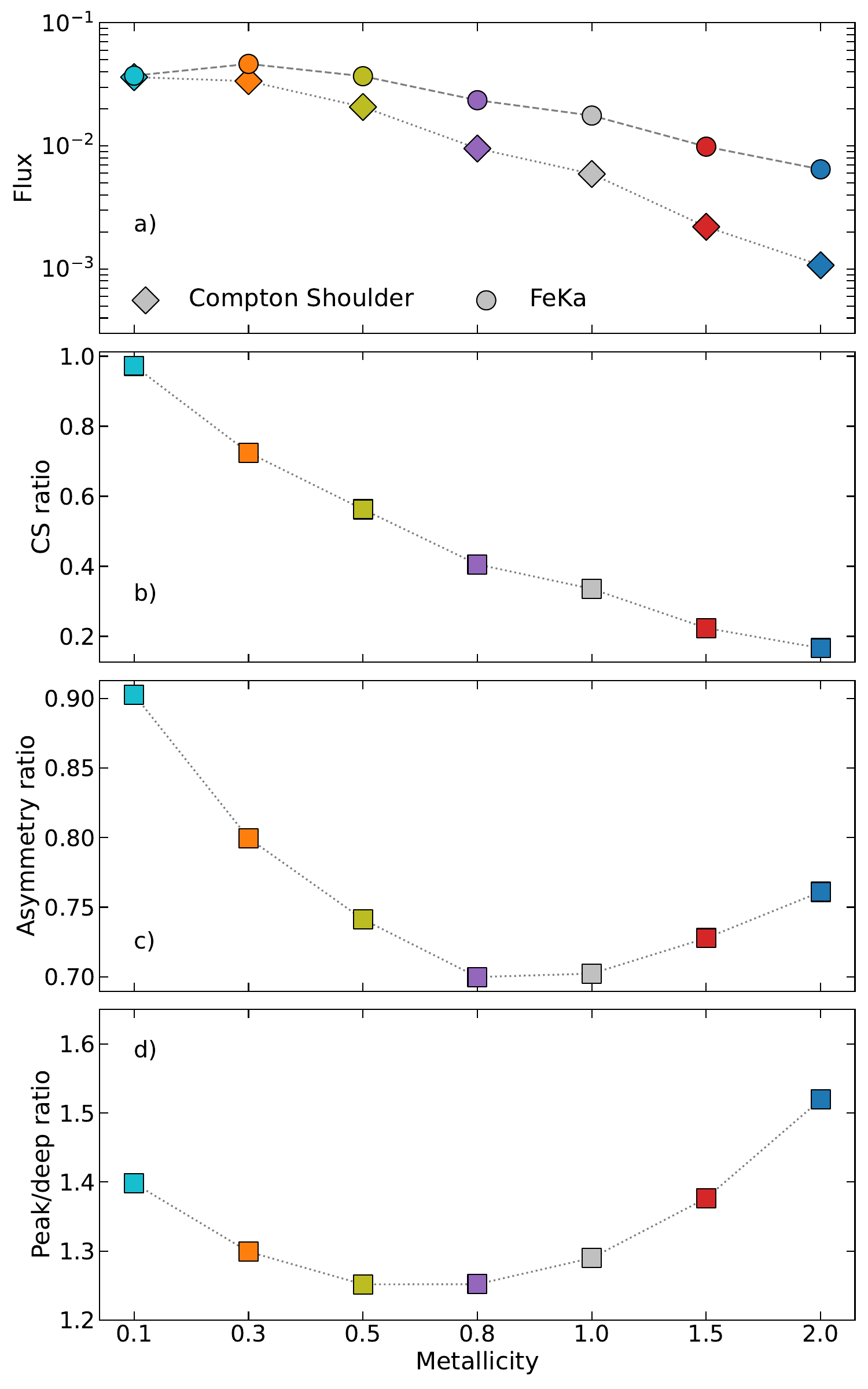}
    \end{subfigure}
    \vspace{-3mm}
    \caption{{\textit{Top}}: Spectrum around the \feka{} line for different metallicities. In all examples we use a spherical corona as X-ray source and a simple toroidal reflector. {\textit{Bottom}}: a) Total flux of the CS and \feka{} b) ratio of the CS flux over the \feka{} line c) asymmetry ratio d) peak/bottom ratio.}
    \label{fig:torusZ}
\end{figure}

In the next set of simulations, we investigate how metallicity ($Z$) affects the Compton shoulder.
As discussed in \S\ref{abnd}, the metals in the torus influence the shape of the CS due to changes in the optical depth of the medium.
To explore this, we run a number of simulations covering a range of metallicities from $Z=0.1-2$, considering the \texttt{lodd} composition, as described in the previous section.
We use the X-ray source and torus described in \S\ref{building_blocks}, with the column density and \h2{} fraction of the torus fixed to the values reported in Table\,\ref{tab:component_details}, $10^{24.5}\,\rm cm^{-2}$ and 0.3 respectively. 
For these simulations, we assume that hydrogen and helium are neutral.

The top panel of Fig.\,\ref{fig:torusZ} displays spectra around the \feka{} line for various metallicities.
Increasing the metallicity leads to a decrease in total flux due to greater absorption, consistent with previous studies (e.g., \citealp{Furui:2016,Odaka:2016}).
Panels (a) and (b) in\,Fig.\,\ref{fig:torusZ} show that the flux of the \feka{} line and the CS both decline as metallicity increases, and their ratio drops accordingly, indicating that the CS is more affected than the \feka{} line.   
The asymmetry and peak/bottom ratio of the shoulder are shown in panels (c) and (d) of Fig.\,\ref{fig:torusZ}.
Both indices follow a negative trend up to $Z=0.8$, without significant in the values of its asymmetry index and an almost stable peak/bottom ratio.
However the rapid increase in absorption that occurs for high metallicities ($Z=1;\,1.5;\,2$) affects significantly the shape of the CS, resulting in a more asymmetric feature.
Our simulations point that both the metal composition and metallicity affect in a similar way the CS.

The effect of metallicity on the CS have been investigated for different geometries by using the MONACO code by \citet{Furui:2016} and \citet{Odaka:2016}. In \citet{Furui:2016}, a toroidal reflector was used with a central X-ray source, considering both a smooth and a clumpy torus.
For the smooth torus, the column density was fixed at $N_{\rm{H}} = 10^{24}\,{\rm cm^{-2}}$ and $10^{25}\,{\rm cm^{-2}}$, while $Z$ varied from 0.1 to 10.
Their results showed that the total flux of the shoulder and the \feka{} line drop as $Z$ increases, while the CS gets a more distinctive shape similar to our findings.
Moreover, they calculated the ratio between the CS and \feka{} flux and found a negative trend with increasing $Z$ similar to the one presented in\,Fig.\,\ref{fig:torusZ} panel (b). \citet{Odaka:2016} performed simulations using a spherical and a slab reflector. 
For the spherical geometry, $N_{\rm{H}}$ was fixed at $5\times 10^{23}\rm\,cm^{-2}$ and $Z$ varied from 0.1 to 10.
The results are similar to \citet{Furui:2016}, showing a decreasing overall flux as the metallicity increases, with a sweet spot at $Z\sim 0.5-1.0$, where the CS shows a prominent peak at $\rm 6.24\,keV$.
Also, they measured the ratio of the equivalent width of the \feka{} over the CS and they found a flat behavior up to $Z \sim 0.5$, followed by a decline for higher metallicities.

\subsubsection{Column density}\label{col_density}

Any means that regulate the optical depths of the intercepting medium affect directly the \feka{} and its CS. 
This can be achieved either by tuning the concentration of the metals in the medium, as discussed in \S\ref{abnd} and \S\ref{metal}, or by altering the total amount of material. To test the second factor, we perform a series of simulations using a torus with varying \nh{}. The rest of the settings are similar to the one used in \S\ref{metal}.
The range of column densities tested is $\log (N_{\rm H}/\rm cm^{-2}) = 23.5 - 25.5$.

\begin{figure}
    \centering
    \begin{subfigure}{\columnwidth}\includegraphics[width=\columnwidth]{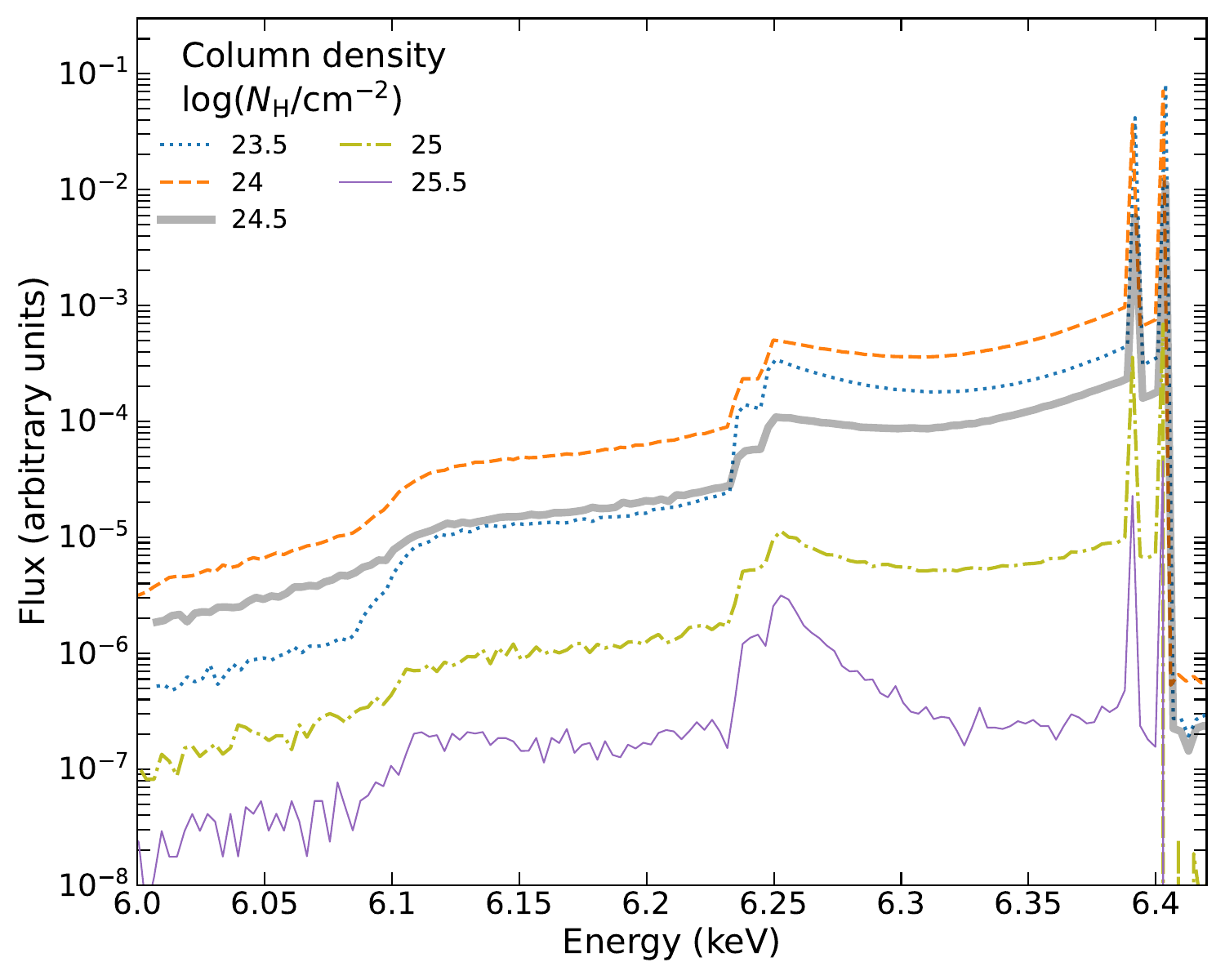}
    \end{subfigure}
    \hfill
    \begin{subfigure}{\columnwidth}\includegraphics[width=\columnwidth]{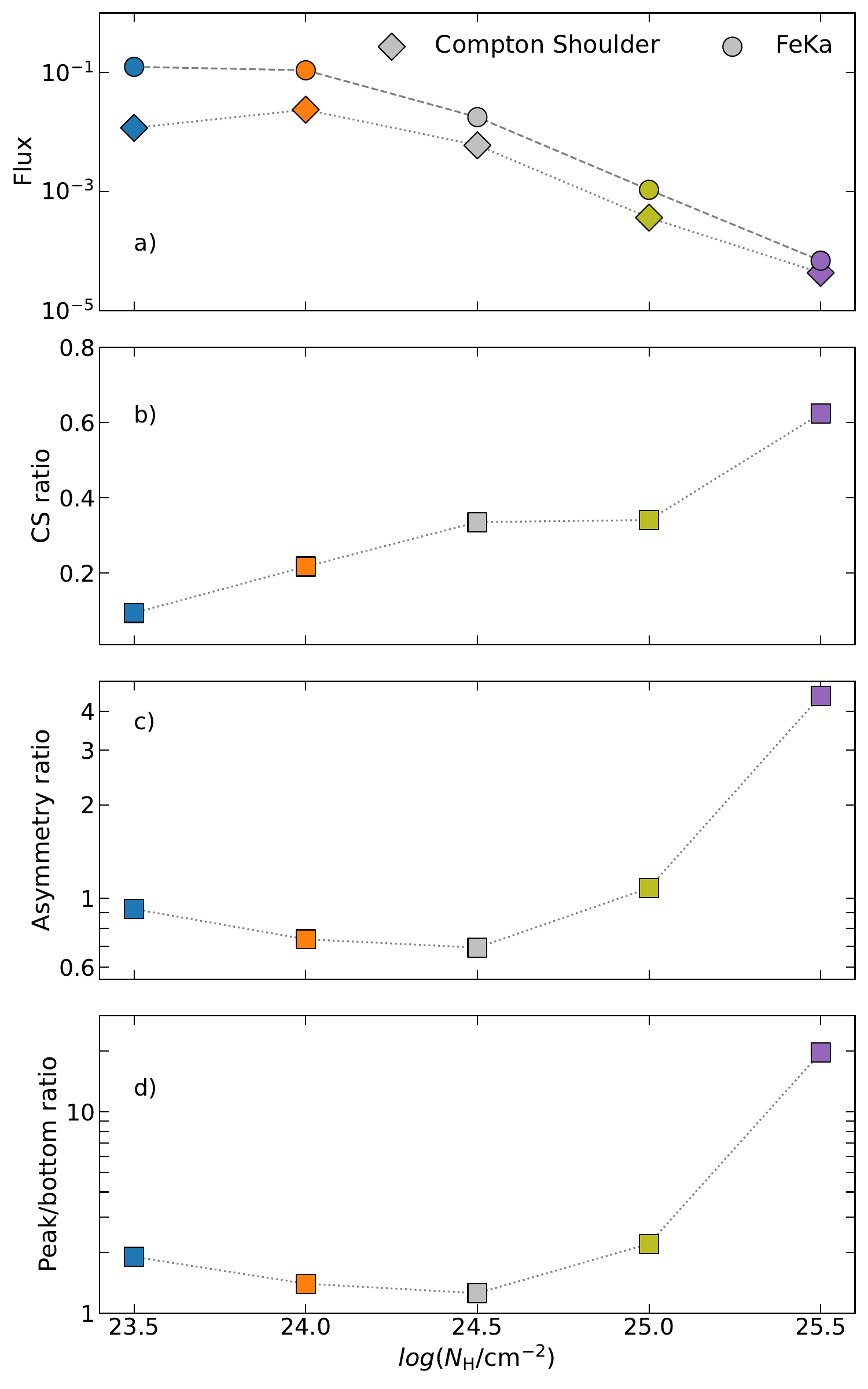}        
    \end{subfigure}
    \vspace{-3mm}
    \caption{{\textit{Top}}: The Compton shoulder for different torus column densities. {\textit{Bottom}}: a) The CS and \feka{} flux, b) The ratio of the CS flux and the \feka{} line, c) The asymmetry ratio and d) peak/bottom ratio.}
    \label{fig:torusNH}
\end{figure}

The spectra of the different simulations are shown in\,Fig.\,\ref{fig:torusNH}-top.
It is illustrated that, for column densities $\log (N_{\rm H}/\rm cm^{-2}) =23.5 - 24$, the flux level is significantly higher than for high column densities [$\log (N_{\rm H}/\rm cm^{-2}) =25 - 25.5$], due to the effect of absorption that gradually increases with \nh{}. 
At high column densities, the shape of the shoulder is characterized by a broad peak around $6.24\rm\,keV$ (back-scattered \feka{} photons).
The behavior of the fluxes and flux ratios of the line and the CS is also explored.
In\,Fig.\,\ref{fig:torusNH}-bottom we present these ratios for the different column densities.
In panel (a) it is shown that the flux of the \feka{} line drops for increasing column densities.
The CS flux shows an increase from $\log (N_{\rm H}/\rm cm^{-2}) = 23.5$ to 24, while the \feka{} flux slightly drops.
As the column density increases both fluxes decline but the \feka{} line shows a steeper decline.
This leads to an increase of the relative ratio between the two features as depicted in panel (b).
The asymmetry of the CS and the peak/bottom ratio show a similar behavior (see panels (c) and (d) in Fig.\,\ref{fig:torusNH}).
For low \nh{} there are no significant changes, but the huge drop in total flux for \nh{} $\geq10^{25}\,cm^{-2}$ causes a rapid increase in both of these ratios.

Other studies have demonstrated a strong correlation between the CS and column density, using different geometries (e.g., \citealp{Matt:2002,Murphy&Yaqoob:2009,Yaqoob&Murphy:2010,Furui:2016,Odaka:2016}). The distinct shape of the CS for toroidal geometries with very high column densities ($N_{\rm H} \geq 10^{25}\,{\rm cm^{-2}}$) when observed edge on has been first reported in \citealt{Yaqoob&Murphy:2010}.
The characteristic back-scattering peak at $\sim 6.24\,{\rm keV}$ is a feature that is a combination of the column density (high column density is necessary for sufficient scatterings), the geometry of the system (since for high column densities most of the CS photons are produced in the inner part of the doughnut shaped medium) and the observing angle which works as proxy of the dominant scattering angle for given \nh{} and geometry.
In the next section we explore in detail this phenomenon.

\subsubsection{Number of scatterings}\label{Sec:scatterings}

\begin{figure}
    \centering
    \begin{subfigure}
        {\columnwidth}\includegraphics[width=\columnwidth]{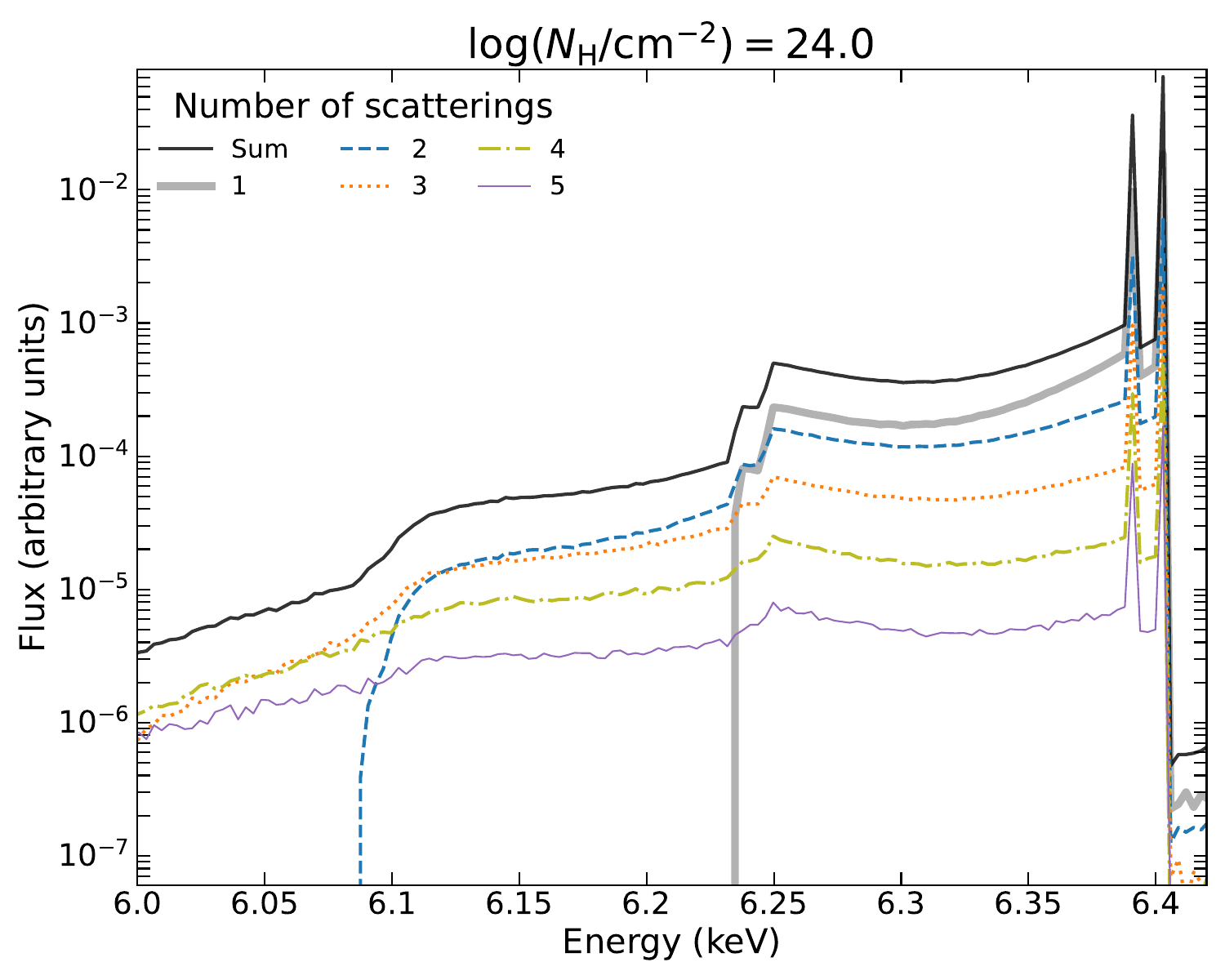}
    \end{subfigure}
    \hfill
    \begin{subfigure}
        {\columnwidth}\includegraphics[width=\columnwidth]{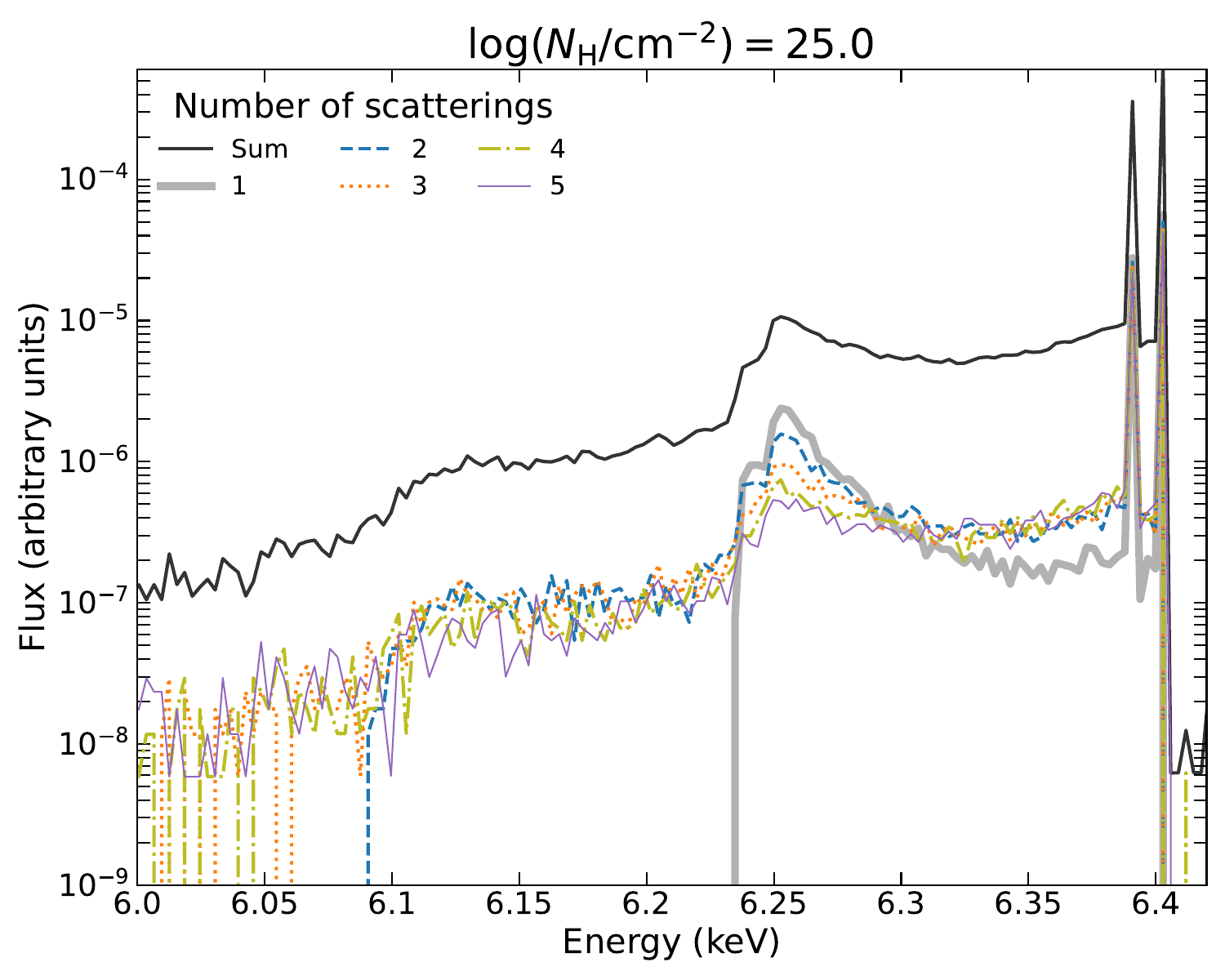}
    \end{subfigure}
    \vspace{-3mm}
    \caption{The \feka{} and CS spectrum formed by fluorescent photons that have undergone a specific number of scatterings (1 to 5). The black line ("sum") is the total spectrum including all the scatterings.
    The model consists of a spherical corona and a torus, with fixed physical properties $Z=1$, \texttt{lodd} in two different column density scenarios.    
    {\textit{Top}}: A torus with column density $N_{\rm{H}}=10^{24.0}\,\rm{cm^{-2}}$. {\textit{Bottom}}: A torus with column density $N_{\rm{H}}=10^{25.0}\,\rm{cm^{-2}}$.}
    \label{fig:scatterings}
\end{figure}

The exact location in the obscuring material responsible for the formation of (most of) the Compton shoulder has been debated.
\citealt{Yaqoob&Murphy:2010} tested a toroidal reflector for various column densities and showed that the CS is dominated by \feka{} photons that have undergone one scattering, which for large observing angles favors the forward and back-scattering.
We ran a series of simulations in which the photons are selected according to the number of scatterings a fluorescent photon undergoes before collected (in \Reflex{} notation: FLUOR $>$ 0, SCATTER == X; X:1-5).
An X-ray source with fixed properties is used (see for details \S\ref{sec:xsource}, along with a torus featuring metallicity $Z=1$ and abundance \texttt{lodd}.
Two cases of torus column density are tested.
One torus with equatorial column density $\log (N_{\rm H}/\rm cm^{-2}) =24.0$ and another more opaque torus with $\log (N_{\rm H}/\rm cm^{-2}) =25.0$.
These two values were selected as examples based on the findings of \S\ref{col_density} where the aforementioned values so significant differences (see,\,Fig.\,\ref{fig:torusNH}).
Both configurations are observed edge on (see, Appendix\,\ref{App:Inclination} for details).
The spectra formed from the given number of scatterings are presented in the two panels of Fig.\,\ref{fig:scatterings} depicted with different colors.
The black solid line with higher flux shows the total flux of the CS and the \feka{} for both scenarios.
The top panel shows the case of $\log (N_{\rm H}/\rm cm^{-2}) =24.0$.
The order of scatterings paves the way to the fluxes and each scattering number gradually contributes less to the CS having less and less flux.
Furthermore, no significant changes in the shape of the CS is present.
This suggests that photons that have undergone a variety of scatterings come from many parts of the torus which means a plethora of scattering angles and therefore more uniform shape.
On the other hand, when the torus becomes even thicker $\log (N_{\rm H}/\rm cm^{-2}) = 25.0$ the view changes.
First, all the orders of scatterings are concentrated in the same flux levels. Moreover, the shape of the CS between the various scattering numbers changes too.
The first scattering (grey-thick line) has a very asymmetric profile depicting the dominance of the back-scattering with a prominent peak at $\sim 6.24$\,keV. The second scattering (blue-dashed line) also has a strong back-scattering peak but in general a smoother CS profile.
Higher order scatterings show a flatter profile once again reflecting the various scattering angles of the photons that form the CS.

\begin{figure*}
    \includegraphics[width=0.95\textwidth]{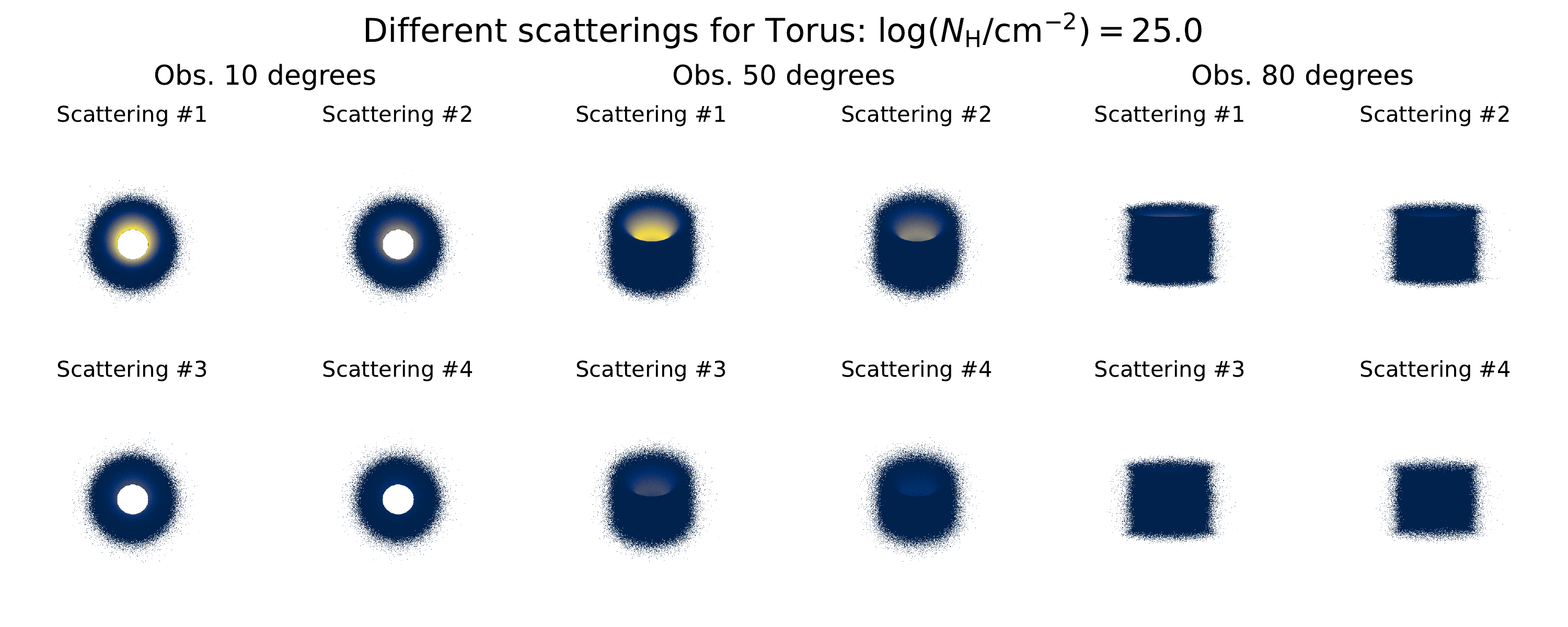}
    \vspace{-3mm}
    \caption{Images generated with \Reflex{} for a torus with $\log (N_{\rm{H}}/{\rm cm^{-2}}) = 25.0$. Each image is created by collecting photons based on the number of scatterings it has undergone before collected. From left to right are presented three different observing angles. The high asymmetry on the Compton shoulder that was found for high \nh{} and low number of scatterings it is illustrated in the yellow core of each doughnut.}
    \label{fig:images_panels_scatterings}
\end{figure*}

As shown in the bottom panel of Fig.\,\ref{fig:scatterings}, for high column density ($N_{\rm H}=10^{25.0}\,{\rm cm^{-2}}$) the shape of the spectrum changes depending on the number of scatterings. We explore the region in the torus where they last scattered before being collected by the detector.
To do that we used the option of \Reflex{} to generate images. We used the exact same configuration as described in Table\,\ref{tab:component_details}, with the source properties fixed to those reported in Table\,\ref{tab:source_details}.
The equatorial column density was set to be $\log (N_{\rm H}/\rm cm^{-2}) = 25.0$.
To see how the CS is formed, we created images by collecting the photons that fall in the $6.20 - 6.42$\,keV energy range (Fig.\,\ref{fig:images_panels_scatterings}). We show the toroidal reflector from three different observing angles, so that the inner regions can be easily observed at low angles ($\sim 10^{\circ}$), while being "obscured" at higher inclination angles ($\sim 80^{\circ}$).
Each panel shows the images that we reconstructed by collecting photons based on the number of scattering from one to four. In the previous paragraph (see, Fig\,\ref{fig:scatterings}) the fifth scattering is included as well but it does not show any significant difference with the fourth one and therefore for clarity we omitted it in this case.
The high asymmetry of the first scattering arises primarily from forward and back scattering due to the energy loss profile associated to Compton scattering. These scatterings take place in the inner surface of the doughnut shape torus (see also \citealt{Yaqoob&Murphy:2010, Furui:2016}). This is illustrated as the brighter central regions observed in the first scattering, especially when viewed from above (10 degrees), where the inner surface is directly exposed. While the second scattering also exhibits some asymmetry with a brighter center, higher-order scatterings show a more uniform light distributions, as shown by the consistent blue color across the images.

\subsubsection{Dust}\label{sec:dust}

The presence of dust in the torus is expected since it extends beyond the sublimation radius, as discussed in \S\ref{sec:torus}. In the most recent version of \Reflex{} \citep{Ricci&Paltani:2023}, the user can adjust the amount of dust particles by changing the \textit{dust depletion} parameter (\textsc{dust}), which controls the fraction of metals that are bound in dust grains. To investigate whether dust affects the shape of the CS, we followed the same approach as in \S\ref{col_density} (see\,Table\,\ref{tab:component_details}) and varied the dust depletion factor to \textsc{dust} 0, 0.7, and 1.0, which correspond to fractions of iron bound in dust grains of 0\%, 70\%, and 100\%, respectively.

\begin{figure}
    \centering
    \includegraphics[width=\columnwidth]{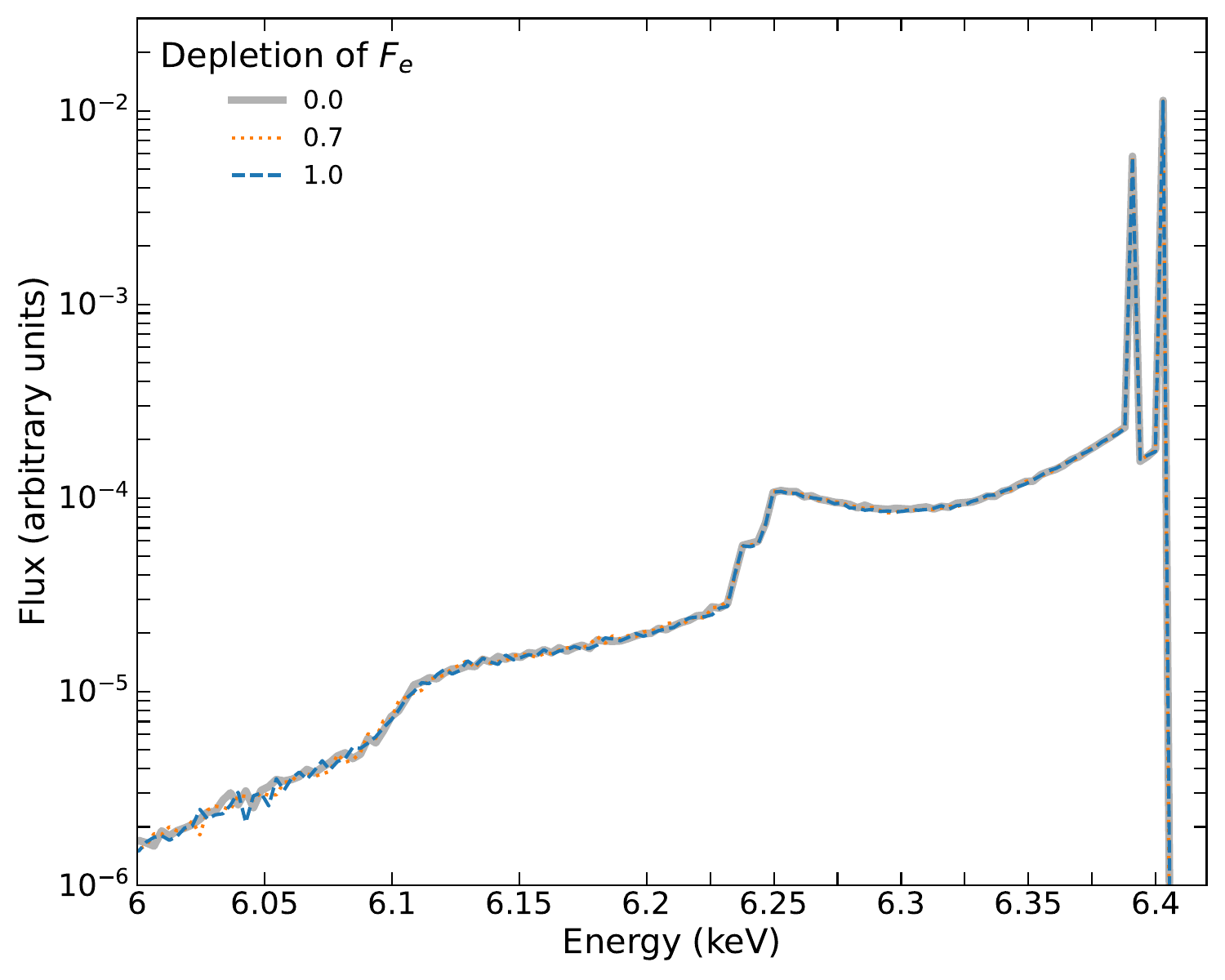}
    \vspace{-3mm}
    \caption{{\textit{Top}}: The Compton shoulder for different values of dust depletion. In all examples we use a spherical corona as X-ray input source along with a torus, observed edge on, with varying dust fractions.}
    \label{fig:torus_dust}
\end{figure}

In\,Fig.\,\ref{fig:torus_dust}-\textit{top} the spectrum at the \feka{} line and CS region is presented. It is clear from the comparison between spectra with different depletion values, that the presence of dust does not affect neither the \feka{} emission line nor its CS. \citet{Ricci&Paltani:2023} tested the effect of dust on fluorescent lines that originate from elements that can be found in dust grains, such as oxygen, carbon, magnesium, silicon and iron. No significant difference in the properties of fluorescent lines was found in their work.

\subsection{Velocity broadening}\label{sec:velbroad}

So far we have investigated configurations of tori that can vary in terms of their physical conditions.
These physical properties can affect the shape of the \feka{} and its CS (i.e. column density) or have zero impact like the presence of dust grains.

However, in the aforementioned simulations we did not take into account potential velocity broadening on the emission line due to various effects such as Keplerian motions of the circumnuclear material.

\begin{figure}
    \centering
    \begin{subfigure}
        {\columnwidth}\includegraphics[width=\columnwidth]{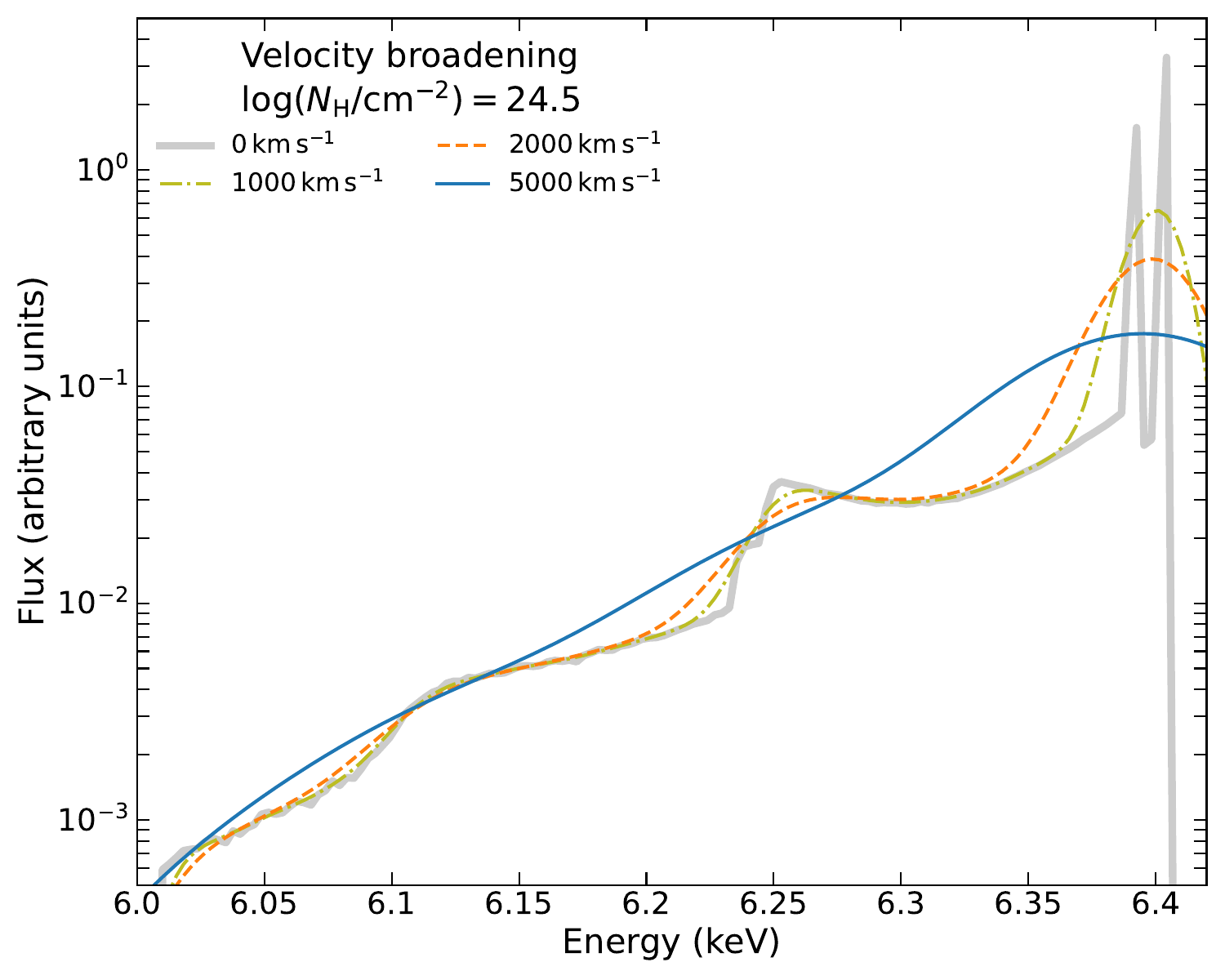}
    \end{subfigure}
    \hfill
    \begin{subfigure}
        {\columnwidth}\includegraphics[width=\columnwidth]{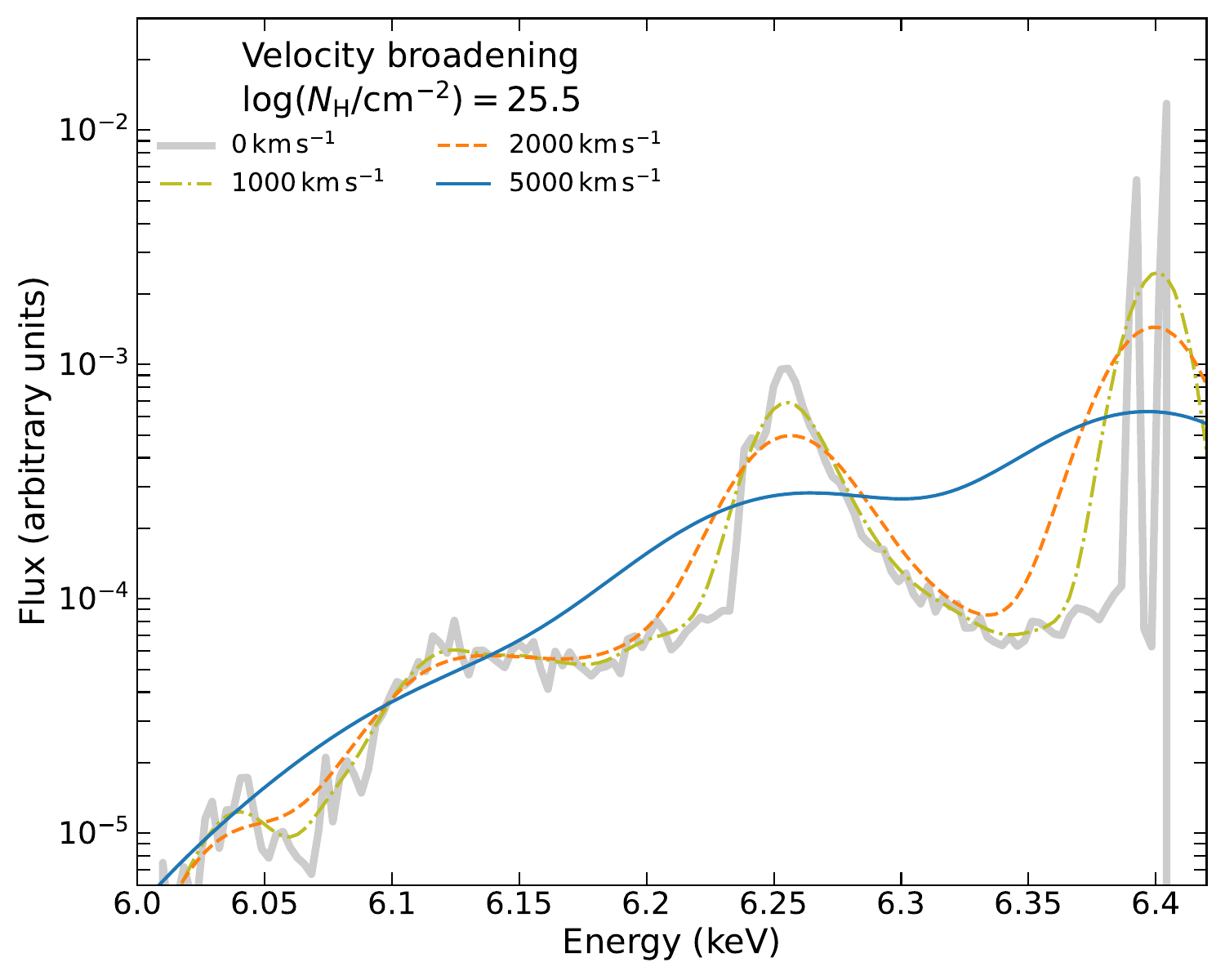}
    \end{subfigure}
    \vspace{-3mm}
    \caption{The \feka{} energy region along with its CS formed by a simple toroidal reflector. \textit{Top}: The panel shows the spectrum for a torus with $N_{\rm H} = 10^{24.5}\, {\rm cm^{-2}}$.
    The different colors represent different gaussian broadening that has been applied to the raw spectral output of the simulation. \textit{Bottom}: Same exercise for a torus with higher column density.
    \label{fig:torus_velocity}}
\end{figure}

In fact, any kind of broadening on the \feka{} complex can affect the resulting spectrum.
Broad \feka{} lines have been observed in several AGN in which the broadening can be attributed to the revolution of chucks of gas at different velocities based on the distance from the central SMBH (e.g., \citealt{Shu+Yaqoob+:2010TypeI,Shu+Yaqoob+:2011TypeII}); an effect that should not be confused with the relativistic broadening that can also be observed which is a result of the general relativity and is often accompanied by distorted soft X-ray spectral features (e.g., \citealt{Brenneman+:2009RelavLines}).
In this section the goal is not to breakdown the different velocity components of the \feka{} or to use the CS to constrain such velocities.
On the contrary we introduce an ad-hoc gaussian broadening to the spectra to emphasize to the fact that the Compton shoulder as a tool to constrain the physical properties of the circumnuclear medium, presents many limitations when the \feka{} complex is affected by any kind of broadening effects.
The version of \Reflex{} that is used throughout this analysis does not include any kind of motions to the material. 
Yet, another version is under development that implements such physical mechanics. 
Nevertheless, available tools in X-rays, such as SKIRT \citep{SKIRT-Xrays:2023}, can be used for a detailed analysis of velocity effects in various media and configurations.
Here, we adopt the same configuration with the previous paragraphs to test the effect of the velocity broadening on the \feka{} and its CS.
A simple torus with its physical properties fixed to the values of Table\,\ref{tab:component_details} is used.
In \S\ref{col_density} we have shown that the column density of the torus can affect the shape of the CS and it can lead to less or more distinctive shapes.
A clear difference is noticed between a torus with $\log (N_{\rm H}/\rm cm^{-2}) > 24.5$ and one with column density less than that, with the first showing higher asymmetry whereas the latter a flatter CS as seen in Fig\,\ref{fig:torusNH}.
Hence we decided to test the velocity broadening effect in two cases. One for $N_{\rm H} = 10^{24.5}\,{\rm cm^{-2}}$ and one for $10^{25.5}\,{\rm cm^{-2}}$.
We apply in the output spectrum of \Reflex{} the gaussian function \texttt{gsmooth}\footnote{https://heasarc.gsfc.nasa.gov/xanadu/xspec/manual/node288.html} from XSPEC to take into account the velocity broadening.
The width of the broadening is:
\begin{equation}\label{eq:vel_broadening}
    \sigma_{\rm E} = \sigma_{\rm L}\left(\frac{E}{6\,{\rm keV}} \right)^{\alpha},
\end{equation}
where
\begin{equation}
    \sigma_{\rm L} = 0.85\left(\frac{V_{\rm FWHM}}{100\, {\rm km/s}} \right)\, {\rm eV}.
\end{equation}
Using \textit{Chandra}/HETG data \citeauthor{Shu+Yaqoob+:2010TypeI}\,(\citeyear{Shu+Yaqoob+:2010TypeI},\,\citeyear{Shu+Yaqoob+:2011TypeII}) studied extensively the \feka{} emission line in several nearby type\,I and type\,II AGN providing tables with the FWHM of the line for each source.
The velocity broadening reported in these works spans from a few thousands to tens of thousands $\rm km\,s^{-1}$.
We have adopted three different values: 1000, 2000 and 5000\,$\rm km\,s^{-1}$ which correspond to $\sigma_{\rm L}= 8.5,\,17,\,42.5\,{\rm eV}$ respectively, for $\alpha=1$.
The different velocities suggest that fluorescence originates in different regions, with higher velocities to be related to the broad line region and the lower ones to material that is even further like the dusty torus.
In Fig.\ref{fig:torus_velocity} the results obtained for the different velocities are presented along with the zero velocity spectrum (grey line).
It is clear that for 1000 and 2000\,$\rm km\,s^{-1}$ the \feka{} line and the CS are still clearly distinguishable.
Although the doublet of the line is diluted and the back-scattering peak of the CS is stretched it still maintains the "shoulder" shape.
In the high-velocity case (i.e. 5000\,$\rm km\,s^{-1}$) the complex of the \feka{} line and the CS are entirely smoothed.
Nevertheless, if we compare the shape of the \feka{} and its CS between the two column densities tested, it is clear that the overall shape is significantly difference. For \nh{} 24.5 the slope is smoother from lower energies to the \feka{} energy whereas for \nh{} 25.5 a bumpy shape is produced. The distinct shape between the two implies that even though the complex (\feka{} + CS) is blended, the overall information for the column density of the material is maintained.
Overall, it is clear that rapidly moving material can limit the potential of the CS as a diagnostic tool for the properties of the circumnuclear material.

\subsection{Geometry of the circumnuclear material}\label{geometries}

In the previous paragraphs we illustrated the effect of the physical properties of the obscuring medium on the shape and flux of the \feka{} line and its Compton shoulder.
Previous studies have been focused on the physical properties of the circumnuclear material using simple geometries and considering only one geometrical object in each simulation.
In this section, we introduce different geometries of the medium surrounding the SMBH, using the option of \Reflex{} that allows the user to add different geometrical objects (see\,\S\ref{reflex}).
In\,Fig.\,\ref{fig:cartoon} all the components that are used to build complex models are presented.

We run simulations with increasing complexity, starting from a simple slab geometry and then moving to more complex AGN models, based on the geometrical objects introduced in \S\ref{building_blocks}.
We fixed the physical properties of the different components to the values reported in Table\,\ref{tab:component_details}.
We consider a spherical corona as X-ray input source and collect the photons that originate from fluorescence.
We begin by testing a simple slab geometry (M1).
To model this we use the accretion disk component with fixed physical properties to the ones presented in Table\,\ref{tab:component_details}.
Next, we test a corona inside a homogeneous spherical medium (M2), to see how the presence of isotropic material can change the shape of the CS.
For the spherical medium we use a sphere with center at the SMBH that extends up to the sublimation zone. This sphere has neutral gas with no dust and column density $N_{\rm H} = 10^{24.5}\,\rm{cm^{-2}}$.
We include in our analysis the torus component (M3) that we used in the previous sections too.
We then consider a more complex model which includes an accretion disk along with the torus (M4). Next, we build a model (M5) that includes an accretion disk, torus and the broad line region (BLR).
The last model (M6) combines the accretion disk, the BLR, the torus and a hollow cone.
For the aforementioned geometries we test two different scenarios regarding the observing angle, one for $45^{\circ}$ and one for $90^{\circ}$.

\begin{figure*}
    \centering
    \begin{subfigure}
        {\columnwidth}\includegraphics[width=\columnwidth]{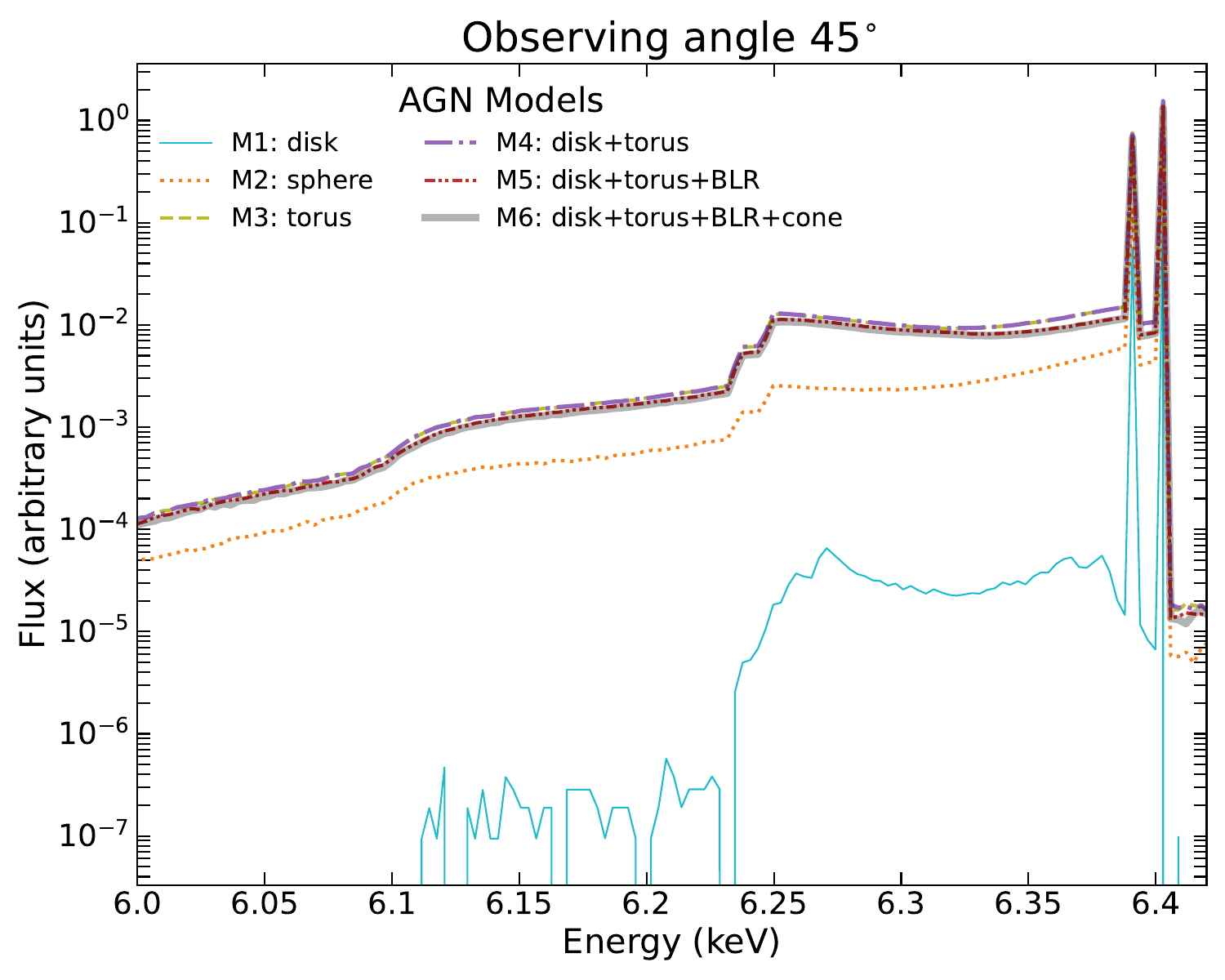}
    \end{subfigure}
    \hspace{0.3cm}
    \begin{subfigure}
        {\columnwidth}\includegraphics[width=\columnwidth]{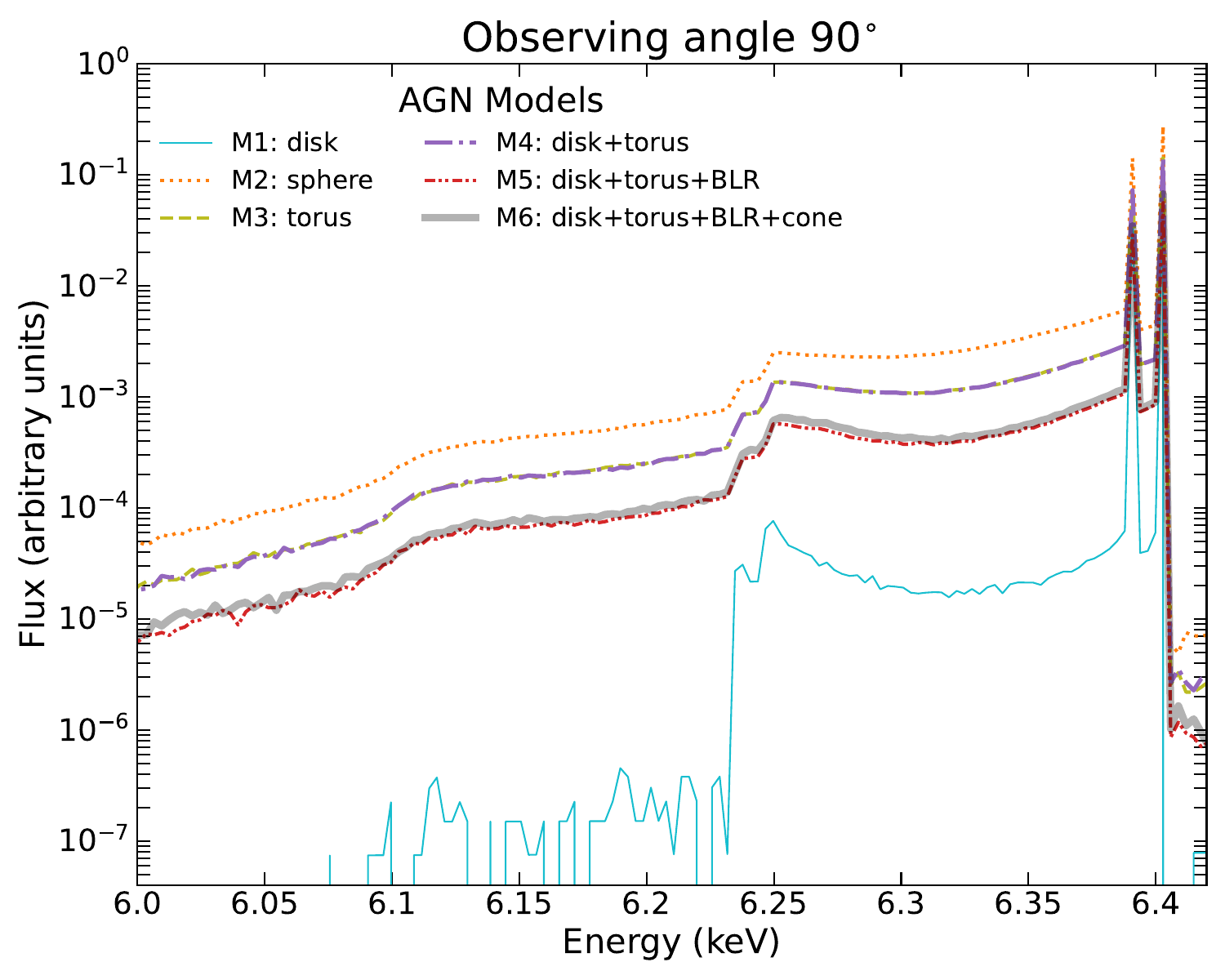}
    \end{subfigure}
    \vspace{0.2cm}
    \begin{subfigure}
        {\columnwidth}\includegraphics[width=\columnwidth]{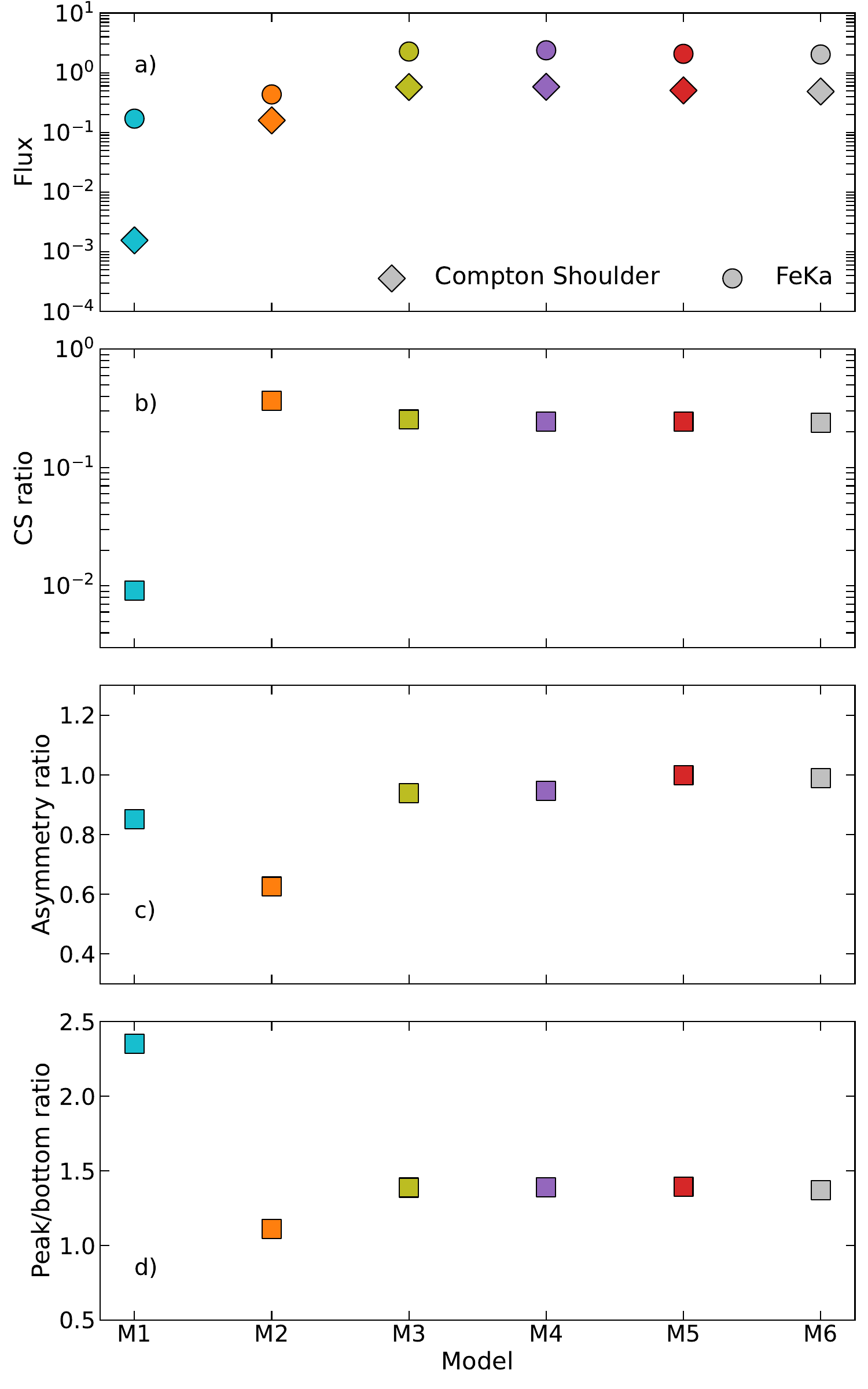}
    \end{subfigure}
    \hspace{0.3cm}
    \begin{subfigure}
        {\columnwidth}\includegraphics[width=\columnwidth]{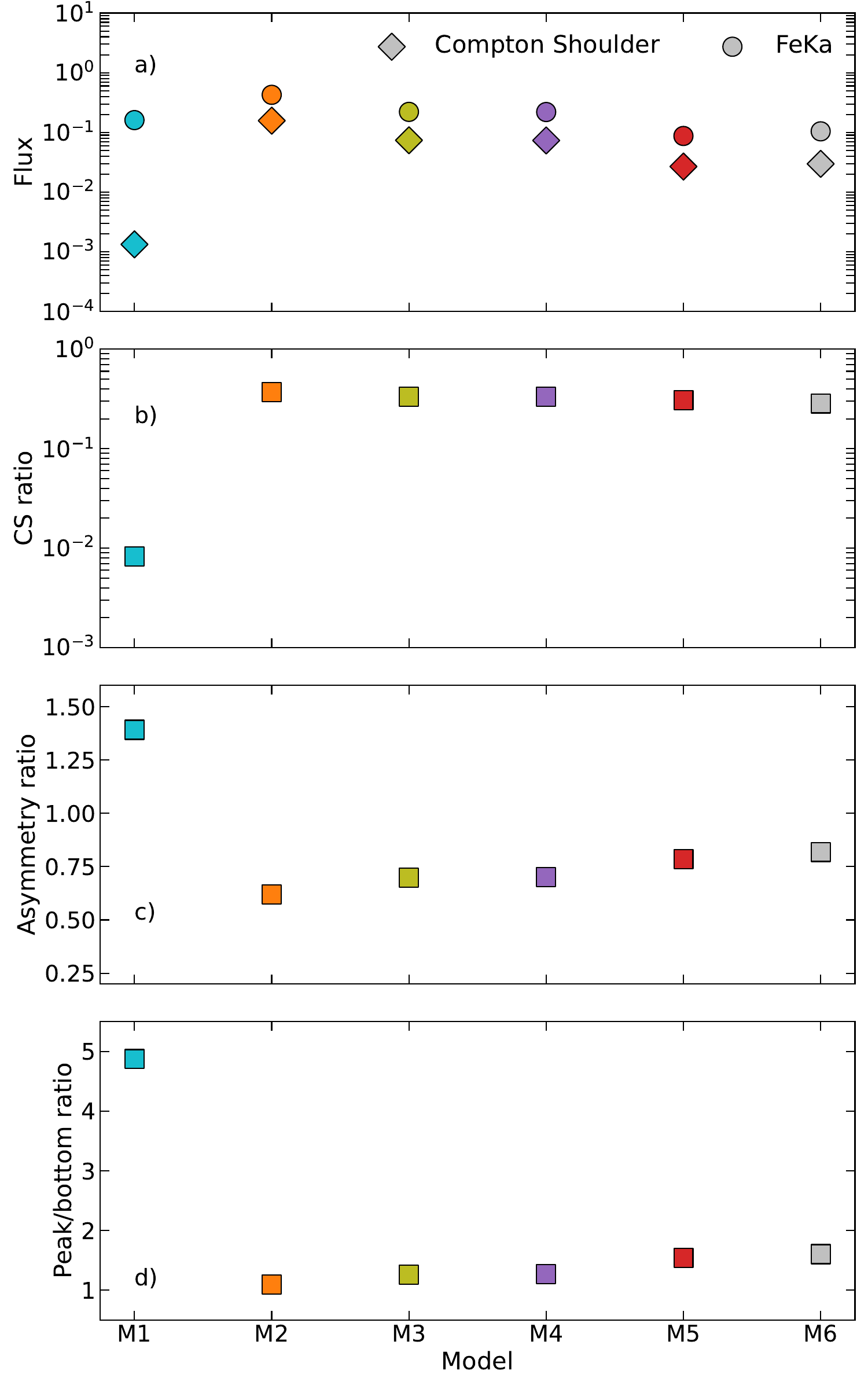}
    \end{subfigure}
    \vspace{-3mm}
    \caption{\textit{Top}: The CS and the \feka{} for different geometries observed in $45^{\circ}$ (left panel) and $90^{\circ}$ (right panel) inclination angle. The models consists of the components presented in\,Fig.\,\ref{fig:cartoon}, with their physical parameters fixed to the values from Table\,\ref{tab:component_details}. The models: M1: \textit{Disk}, M2: \textit{Sphere}, M3: \textit{Torus}, M4: \textit{Disk} + \textit{Torus}, M5: \textit{Disk} + \textit{Torus} + \textit{BLR}, M6: \textit{Disk} + \textit{Torus} + \textit{BLR} + \textit{Hollow Cone}. \textit{Bottom}: a) The CS and \feka{} flux, b) The ratio of the CS flux and the \feka{} line, c) The asymmetry ratio and d) peak/bottom ratio.}
    \label{fig:geometries}    
\end{figure*}

The spectra focused on the \feka{} line region are illustrated in\,Fig.\,\ref{fig:geometries}.
The top panels show the spectra, whereas at the bottom panels the fluxes and ratios can be found.
First, for the case of the disk geometry (M1) we can see that the flux level for both observing angles is similar.
The shape of the CS, however, shows a different peak based on the scattering angle.
Different observing angles favor the collection of photons that have been scattered for different angles as shown in details in \citealt{Yaqoob&Murphy:2010}.
In Appendix \ref{App:Inclination}, we further test the effect of the observing angle on the CS in the case of slab geometry (see Fig.\,\ref{fig:slab_angle}.
For the case of the spherical surrounding material (M2) the flux of the CS is consistent for both angles since it is a fully symmetric configuration.
As seen in Fig.\ref{fig:geometries} panel (c) the spherical CS is smoother or less asymmetric another effect the isotropic nature of the spherical geometry.
Models M3, M4, M5 and M6 include the torus, which is the main medium that reprocess the photons in the edge-on regime and introduces asymmetry to the system. The CS is almost indistinguishable in the case of $45^{\circ}$ angle, something that can be confirmed by the bottom panels (a, b, c, d) of the left panel in\,Fig.\,\ref{fig:geometries}.
When the system is observed edge on ($90^{\circ}$), the presence of the BLR (M5, M6) adds as an extra layer of material between the observer and the source and hence, the flux of the CS and the \feka{} drops.
The hollow cone that is included in M6 does not affect the CS significantly as a result of the low column density of the polar medium.

Overall, it is clear that the slab reflector is a special case of reflecting material something that is also seen in Fig.\,\ref{fig:slab_angle}.
Moreover, the spherical geometry can also be distinguished due to its symmetrical shape with no favorable scattering angle.
Last, all the models that introduce axisymmetry (simple torus M3, torus+disk M4, torus+disk+BLR M5 and and torus-disk-BLR-hollow cone M6) have similar results and it is difficult to distinguish between geometries.
On the contrary the main factor that affects the shape of the CS is the total line-of-sight column density.

\section{Simulated XRISM and Athena spectra}\label{sec:simulations}

The Compton shoulder could be a useful tool to constrain some of the properties of the circumnuclear material. Current X-ray facilities have limited resolution in the $\rm 6-7\,keV$ band and so far the CS has been detected only in a handful of nearby objects (e.g., \citealt{Sato:+1986,Watanabe+:2003,Torrejon:2010,Hikitani:2018}).
Future instruments like \textit{XRISM}/Resolve \citep{xrism:2020} and \textit{Athena}/X-IFU \citep{AthenaXIFU} will have the necessary spectral resolution ($\rm 5-7\,eV$ and $\rm 3\,eV$ for Resolve and X-IFU, respectively) and sensitivity to make the \feka{} CS a useful observable spectral feature.

We used \Reflex{} to create a realistic model of the Circinus galaxy, since it is one of the X-ray brightest nearby heavily obscured AGN, and it is known to show a \feka{} CS feature \citep{Bianchi:2002,Hikitani:2018}.
Since the CS has not been observed in other AGN and as a result it cannot be considered a diagnostic tool yet, it is reasonable to start from a source with an observed CS. Nevertheless, this does not guarantees that in other sources the CS can be effective.
We ran a simulation using the model M6, which consists of an accretion disk, BLR, dusty torus and polar medium (see\,\S\ref{geometries}), a schematic of which can be found in\,Fig.\,\ref{fig:cartoon}.
The physical properties of the different components of the model are fixed to the values reported in Table\,\ref{tab:component_details}.
The column density of the torus is set to $N_{\rm H}= 10^{24.5}\,{\rm cm^{-2}}$ and the metallicity $Z=1$, values that are close to the ones reported in recent studies of the Circinus (e.g., \citealt{Andonie+:2022}). 
The flux level used is the observed $\rm 2-10\,keV$ flux of Circinus reported by \cite{Ricci+:2017Catalog} and the observing angle is fixed to $78^{\circ}$ (\citealt{Arevalo+:2014,Andonie+:2022}).
Velocity broadening around the \feka{} line has been reported in Circinus, with average value estimated to be ${\rm FWHM}=1346\,{\rm km\,s^{-1}}$  \citep{Andonie+:2022}. 
To include the effect of velocity fields, we used \texttt{gsmooth} from XSPEC similarly to \S\ref{sec:velbroad} setting $\sigma_{\rm L}=11.44\,{\rm eV}$ and $\alpha=1$, which corresponds to the FWHM above.

To generate the simulated spectra the X-ray analysis package XSPEC was used along with the response files of the Resolve\footnote{\scriptsize{https://heasarc.gsfc.nasa.gov/FTP/xrism/prelaunch/simulation/sim3/}} and X-IFU\footnote{\scriptsize{http://x-ifu.irap.omp.eu/resources/for-the-community}} instruments, including their respective background files.
The exposure time for the simulated spectra was set to 200\,ks, which provides sufficient signal-to-noise to reveal the CS feature.
For the analysis and the fitting of the simulated data we used XSPEC too.
The spectra are rebinned so there is at least one data point in each energy bin (\texttt{group min 1}). Cash statistics \citep{Cash:1979} are applied and the errors when constrained successfully were calculated in 90\% confidence level throughout our analysis.
After simulating spectra for both instruments, we conducted two tests to evaluate the effectiveness of the CS, focusing on the energy range $6.0-8.0\,{\rm keV}$, which is expected to be dominated by reprocessed radiation for the Circinus galaxy. First, we test whether the input parameters of the simulated spectrum can be retrieved, such as $\Gamma$, $\sigma_{\rm L}$ (FWHM), \nh{} and $Z$.
Then, we applied three different geometrical models from \S\ref{geometries} - torus, sphere, M6 model - to investigate whether the \feka{} and its CS as a spectral feature can differentiate between the given geometries.

\subsection{Testing the physical properties of the torus}\label{sec:simulation_properties}

We develop a grid spectral model considering the M6 geometry, which was introduced in \S\ref{geometries} and includes all the main component of AGN (disk, BLR, torus, polar cone).
Our grid model has been implemented having three free parameters, the spectral index ($\Gamma$), the column density (\nh{}) and metallicity ($Z$) of the torus.
The properties of all the other components are set to the values presented in Table\,\ref{tab:source_details} and \ref{tab:component_details}.
The primary continuum spectral index was set with three different values, 1.5, 1.8, 2.4. Similarly, metallicity was set to values: $Z = 0.5,\ 1.0,\ 1.5$.
The \nh{} spans between $\log (N_{\rm H}/\rm cm^{-2}) = 24\ \rm{and}\ 25$ with steps of $\Delta \log (N_{\rm H}/\rm cm^{-2})=0.1$.

\begin{figure*}
     \includegraphics[width=0.95\textwidth]{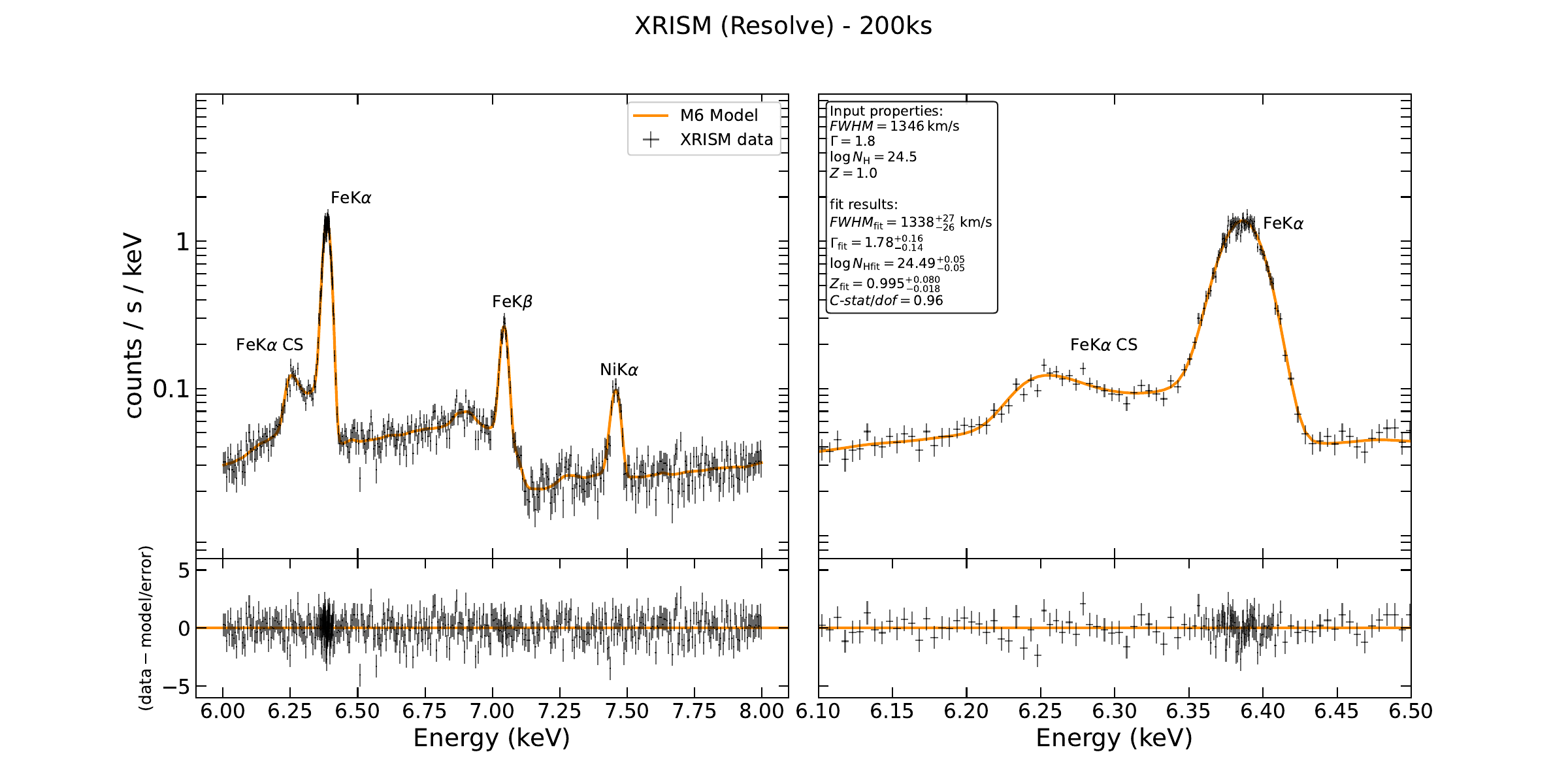}
    \vspace{-3mm}
    \caption{The simulated spectra of Circinus around the \feka{} line region.
    The spectra are fitted with a grid model developed using the M6 geometry (see\,\S\ref{geometries}) Free parameters are the \nh{} and $Z$. In the left panel the \textit{XRISM}/Resolve data along with the data/model ratio are presented. The orange line is the best fit with the retrieved values summarised in the box.}
    \label{fig:sim_xrism_properties}
\end{figure*}

We performed simulation generating photons from $5$ to $50\,keV$, collecting the $6-8$ keV energy range with 3 eV binning, so as to achieve good signal to noise in reasonable computational times.
Having developed the necessary grid output we compiled them in one table model and imported in XSPEC.
We employed this grid model to analyze the simulated spectra from \textit{XRISM} and \textit{Athena} to determine whether the input properties can be retrieved.
We fit the data, in the $6-8$ keV band, using \texttt{gsmooth}\,(\texttt{M6model}) so as to take into account the velocity broadening.
We fit the Circinus \textit{XRISM}/Resolve simulated spectrum and we get $\rm{FWHM}=1338^{+27}_{-26}\,{\rm km/s}$, $\Gamma=1.78^{+0.16}_{-0.14}$, $\log (N_{\rm{H}}/{\rm cm^{-2}}) = 24.49^{+0.05}_{-0.05}$ and metallicity $Z = 0.995^{+0.080}_{-0.018}$.
The model is able to well fit the data ($C$-$stat/dof = 0.96$). 
In the left panel of Fig.\,\ref{fig:sim_xrism_properties} we illustrate the simulated spectrum in the 6-8 keV band along with the best fit. The bottom panel shows the residuals. The right panel of Fig.\,\ref{fig:sim_xrism_properties} is focused on the \feka{} line energy region.
The model provides a good fit and succeeds in constraining the spectral index accurately. For the remaining three properties FWHM, \nh{} and $Z$ it returns values marginally lower than the input ones, even if we consider the errors which are small.
This can be attributed to systematic errors that have not been taken into account.

\begin{figure*}
    \includegraphics[width=0.95\textwidth]{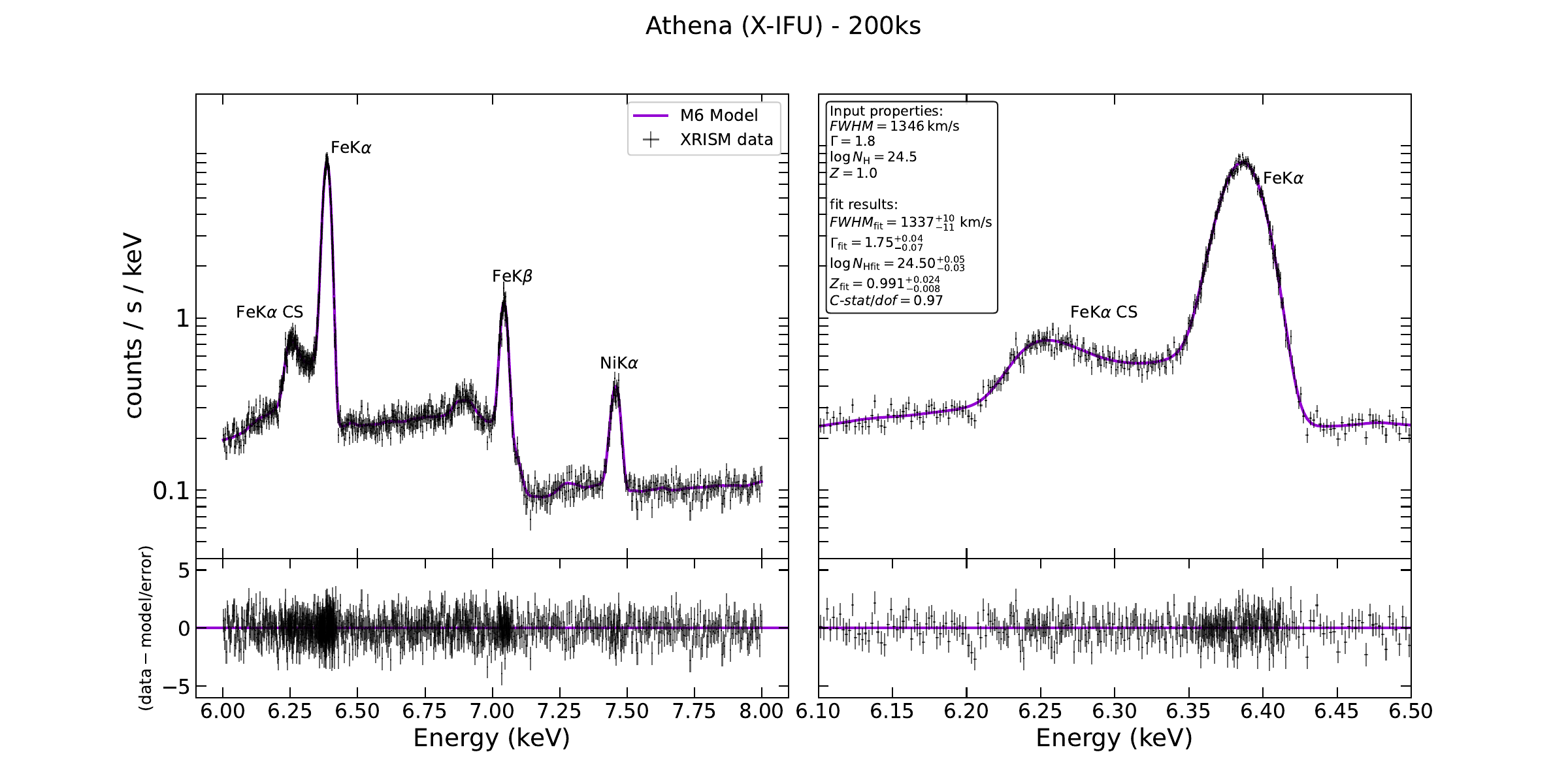}
    \vspace{-3mm}
    \caption{The simulated spectra of Circinus around the \feka{} line region.
    The M6 geometry (see\,\S\ref{geometries}) has been used to generate a grid model with having free parameters the \nh{} and $Z$. In the left panel the \textit{Athena}/X-IFU data along with the data/model ratio are presented. The orange line is the best fit with the retrieved values summarised in the box.}
    \label{fig:sim_athena_properties}
\end{figure*}

Next, we apply the same model on the \textit{Athena}/X-IFU spectrum (see Fig.\,\ref{fig:sim_athena_properties}).
In this case the fit is also good ($C$-$stat/dof = 0.97$) similar to the $XRISM$ example.
The constrained values are: $\rm{FWHM}=1337^{+10}_{-11}\,{\rm km/s}$, $\Gamma=1.75^{+0.04}_{-0.07}$. The column density is $\log (N_{\rm{H}}/{\rm cm^{-2}}) = 24.50^{+0.05}_{-0.03}$ whereas the metallicity is found $Z = 0.991^{+0.024}_{-0.008}$.

Overall, we see that, with the advent of new X-ray facilities, modelling of the \feka{} region can contribute to the understanding of the physical conditions of the circumnuclear material.
In our example the simulated spectra and the model used to fit that are the same and therefore the result cannot be considered as robust evidence for the derived values of the properties. However we consider it as a useful exercice to illustrate how the $6-8\,keV$ band will look with microcalorimeters, as well as explore some profound degeneracies between the properties.

\subsection{Testing different geometries}\label{sec:simulation_geom}

It has been illustrated that the physical properties can be retrieved by fitting the CS even with some deviations from the input values. Despite that, we are interested in testing whether the CS can be used as a tool to distinguish between different geometries. In \S\ref{geometries} several configurations have been tested.
We adopt the following three geometrical models and run simulations using \Reflex{}.
The models are: i) sphere (M2), ii) torus (M3), and iii) M6 (disk, BLR, torus, polar cone) from \S\ref{geometries}, as they show variations in the shape of the CS (see\,Fig.\,\ref{fig:geometries}).
The three grid models have the same free parameters; \nh{}, $Z$ and the spectral index $\Gamma$.
The \nh{} covers $\log (N_{\rm H}/\rm cm^{-2}) = 24\ \rm{to}\ 25$ with steps of $\Delta \log (N_{\rm H}/\rm cm^{-2})=0.1$. For the metallicity and spectral index the same grid with the previous section was selected; $Z = 0.5,\ 1.0,\ 1.5$ and $\Gamma=$ 1.5, 1.8, 2.4 respectively.

\begin{figure*}
    \includegraphics[width=0.95\textwidth]{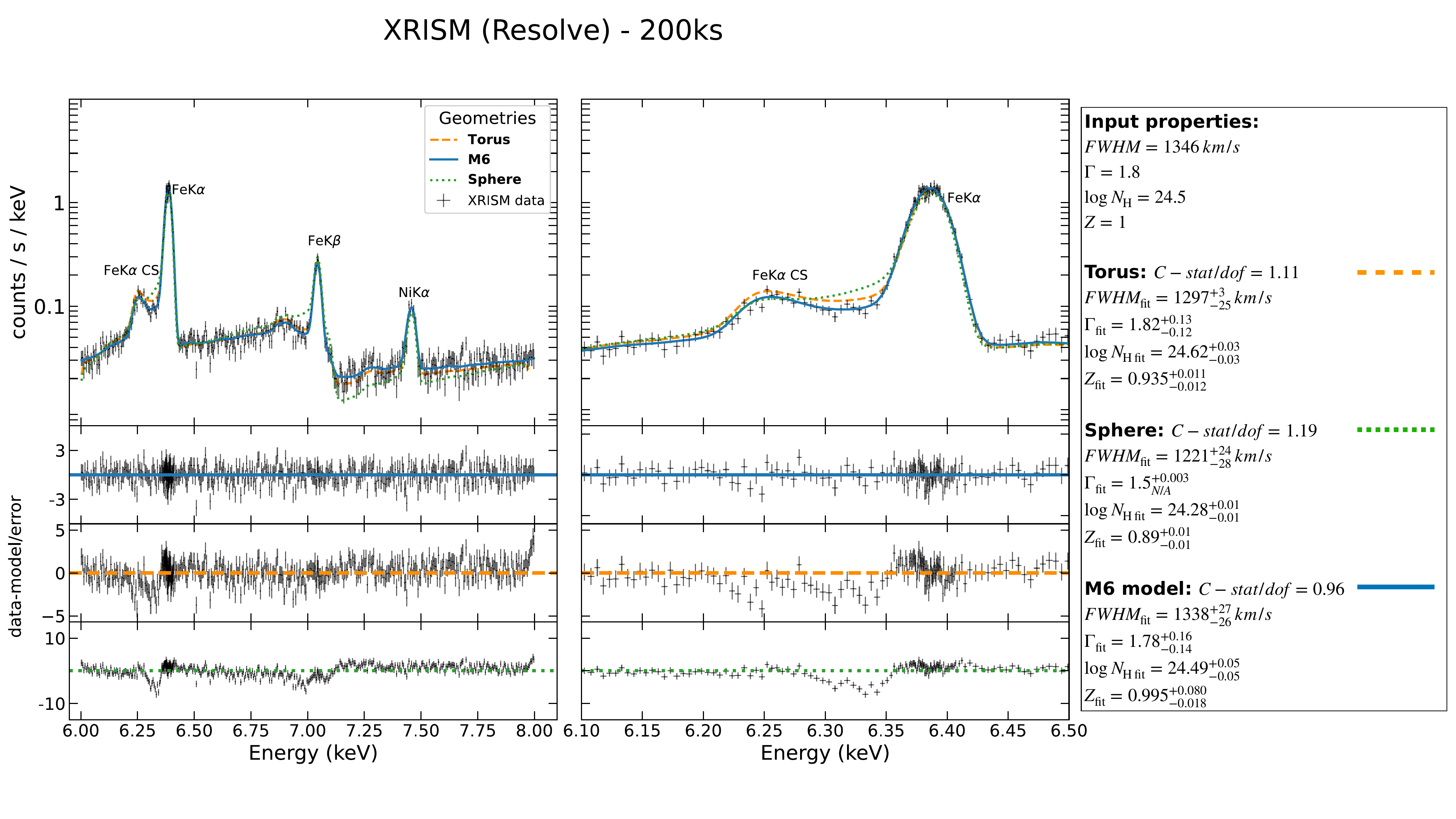}
    \vspace{-3mm}
    \caption{The simulated spectra of Circinus around the \feka{} line region.
    The spectra are fitted with a grid model developed using the M6 geometry (see\,\S\ref{geometries}) having free parameters the \nh{} and $Z$. In the left panel the \textit{XRISM}/Resolve data along with the data/model ratio are presented. The orange line is the best fit with the retrieved values summarised in the box.}
    \label{fig:sim_xrism_geometris}
\end{figure*}

All three models have been applied to both the \textit{XRISM} and \textit{Athena} simulated spectra.
In Fig.\,\ref{fig:sim_xrism_geometris} the simulated spectra along with the best fits and the residuals are presented.

First, for the \textit{XRISM}/Resolve simulated data, when the simple toroidal model is used, the fit shows good statistics with $C$-$stat/dof=1.11$. The values of the input properties retrieved are: $\rm{FWHM}=1297^{+3}_{-25}\,{\rm km/s}$, $\Gamma=1.82^{+0.13}_{-0.12}$, $\log (N_{\rm{H}}/{\rm cm^{-2}}) = 24.62^{+0.03}_{-0.03}$ and $Z=0.935^{+0.011}_{-0.012}$.
It is clear that the toroidal models underestimates the velocity broadening of the line while the column density is slightly larger than the input value, which is a consequence of the BLR that provides an extra layer of absorption.
The above suggest potential degeneracies between the properties.
.

For the spherical model (green line of Fig.\,\ref{fig:sim_xrism_geometris}) the fit gives a relatively accepted $C$-$stat/dof=1.19$.
However, the values it returns diverge from the input ones.
The velocity broadening is found $\rm{FWHM}=1221^{+24}_{-28}\,{\rm km/s}$.
It fails to constrain successfully the photon index $\Gamma=1.5^{+0.003}_{N/A}$. Then the spherical medium was found to have $\log (N_{\rm{H}}/{\rm cm^{-2}}) = 24.28^{+0.01}_{-0.01}$ and $Z=0.89^{+0.01}_{-0.01}$

The M6 model (blue line of Fig.\,\ref{fig:sim_xrism_geometris}) is the same used in the previous paragraph and therefore the results are summarized as:
$C$-$stat/dof=0.96$; $\rm{FWHM}=1338^{+27}_{-26}\,{\rm km/s}$; $\Gamma=1.78^{+0.16}_{-0.14}$; $\log (N_{\rm{H}}/{\rm cm^{-2}}) = 24.49^{+0.05}_{-0.05}$; $Z=0.995^{+0.080}_{-0.018}$).

Overall, we find that the simple torus models succeeds in constraining $\Gamma$ and the properties of the torus considering the lack of BLR in this case, but underestimates the broadening.
The spherical model, as expected from \S\ref{geometries} performs the worst can be ruled out when compared to axisymmetrical geometries like the toroidal models (only torus and full M6 model).
While the values of the uncertainties reported here are typically low, it should be taken into account that there are inherent systematic uncertainties associated to our lack of knowledge of the exact geometrical configuration of the circumnuclear material.

\begin{figure*}
    \includegraphics[width=0.95\textwidth]{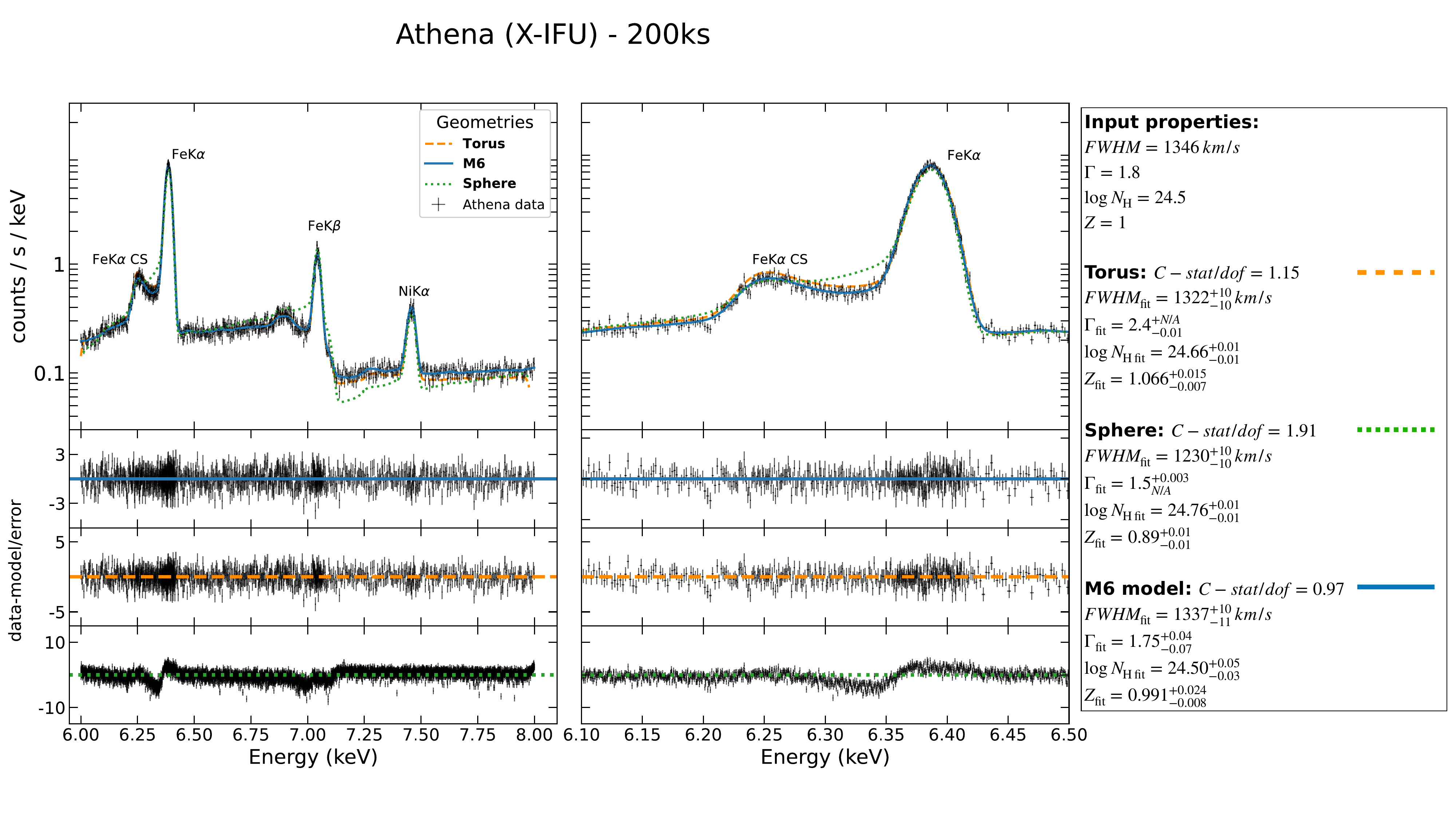}
    \vspace{-3mm}
    \caption{The simulated spectra of Circinus around the \feka{} line region.
    The spectra are fitted with a grid model developed using the M6 geometry (see\,\S\ref{geometries}) having free parameters the \nh{} and $Z$. In the left panel the \textit{Athena}/X-IFU data along with the data/model ratio are presented. The orange line is the best fit with the retrieved values summarised in the box.}
    \label{fig:sim_athena_geometris}
\end{figure*}

In the case of the \textit{Athena}/X-IFU data, the same behavior is reported (see Fig.\ref{fig:sim_athena_geometris}), with the spherical models failing to well reproduce the data. The torus model succeeds in constraining the FWHM and as expected overestimates the \nh due to the lack of the BLR. The values derived are: $\rm{FWHM}=1322^{+10}_{-10}\,{\rm km/s}$, $\Gamma=2.4^{N/A}_{-0.01}$, $\log (N_{\rm{H}}/{\rm cm^{-2}}) = 24.66^{+0.01}_{-0.01}$ and $Z=1.066^{+0.015}_{-0.007}$ for $C$-$stat/dof=1.15$.
For the sphere we found: $C$-$stat/dof=1.91$
$\rm{FWHM}=1230^{+10}_{-10}\,{\rm km/s}$.
It fails to constrain the photon index $\Gamma=1.5^{+0.003}_{N/A}$. Then the spherical medium was found to have $\log (N_{\rm{H}}/{\rm cm^{-2}}) = 24.76^{+0.01}_{-0.01}$ and $Z=0.89^{+0.01}_{-0.01}$.

Finally, the M6 model returns the velocity broadening $\rm{FWHM}=1337^{+10}_{-11}\,{\rm km/s}$, the photon index
$\Gamma=1.75^{+0.04}_{-0.07}$, while the column density is $\log (N_{\rm{H}}/{\rm cm^{-2}}) = 24.50^{+0.05}_{-0.04}$ and $Z=0.991^{+0.024}_{-0.008}$, for $C$-$stat/dof=0.96$.

To conclude, we see that with these new, high-resolution data, the spherical geometry (M2 model) can be easily distinguished from the toroidal (M3) and complex AGN model (M6).
Nevertheless, the latter two models (M3-torus, M6-complex) show degeneracies between the FWHM, the spectral index and the column density. 
This finding suggests that the use of the energy range $6-8\,keV$, including the \feka{} and its CS, has limitation in helping constrain the properties of the circumnuclear material.
 
\section{Summary}\label{summary}
The Compton shoulder is a feature formed by fluorescent \feka{} photons that are Compton-scattered to lower energies by the circumnuclear material around the SMBH.
Since the CS is result of Compton scattering its shape and flux depends on the physical properties of the medium as well as on the geometry of the system.
We have set up a series of simulations, using the ray-tracing platform \Reflex{}, featuring different physical properties as well as geometries to explore their impact on the \feka{} CS (for details see \S\ref{sim_setup}).
In our main analysis we were focused on the properties of the circumnuclear medium and we decided to keep the inclination angle of the system fixed to the edge-on case ($85^{\circ}-90^{\circ}$).
The observing angle is not always known but we consider it an effect related to the orientation between the source and the observer and therefore not a physical property of the medium.
However, it can indeed affect the spectral shape and therefore in Appendix\,\ref{App:Inclination} we explored briefly the effect of it on the \feka{} and its CS.

In \S\ref{sec:physical_prop}, we present the influence of different physical properties on the CS, whereas in \S\ref{geometries} we use different geometries to make complex AGN models with their physical properties fixed.  
We summarize here our results:
\begin{itemize}
\item Metal composition: In \S\ref{abnd} we tested the three more commonly-used abundance tables that can be found in XSPEC, \texttt{angr}, \texttt{wilm}, \texttt{lodd}.

We found that the behavior of the CS is affected by absorption that is regulated by the use of various abundance tables. Table such as \texttt{angr} results in a more prominent shoulder, but the total spectrum suffers from high absorption.
On the other hand, when we used \texttt{wilm}, \texttt{lodd}, which have similar abundance in iron, a less absorbed spectrum with less asymmetrical and thus prominent CS is observed (see\,Fig.\,\ref{fig:torus_abnd}).
\item Metallicity: In \S\ref{metal} the metallicity ($Z$) was tested in order to explore whether it can affect the shape and the flux of the \feka{} and its CS.
Different values of $Z$ regulate the metals and therefore the absorption of the incident photon field.
We tested $Z$ from $0.1-2$.
In\,Fig.\,\ref{fig:torusZ} as the $Z$ increases, the total flux of the spectrum drops.
The flux of the CS also decreases with the metallicity. At high $Z$ ($1.5,\ 2$) the shoulder becomes asymmetrical with high peak/bottom ratio (see panels (c), (d) in\,Fig.\,\ref{fig:torusZ}).
Our results are in agreement with previous works (e.g., \citealt{Furui:2016,Odaka:2016}).
\item Column density: In \S\ref{col_density} we tested different values of \nh{} ($N_{\rm{H}}=10^{23.5} - 10^{25.5}\,{\rm{cm^{-2}}}$).
The spectra for the different column densities are presented in\,Fig.\,\ref{fig:torusNH}.
Optically thin material is incapable of forming a distinctive CS.
On the other hand, high column densities lead to higher absorption, which significantly decreases the overall flux of the CS.
The CS shows a characteristic peak at $\rm 6.24\, keV$ and becomes 
more asymmetrical at $N_{\rm H} \geq 10^{25}\,{\rm cm^{-2}}$ as shown in 
in panels (c) and (d) of Fig.\,\ref{fig:torusNH}.
Previous studies have presented a similar behavior of the CS with varying \nh{} (see, \citealt{Yaqoob&Murphy:2010}, \citealt{Furui:2016}, \citealt{Odaka:2016}).
\item Number of scattering: In \S\ref{Sec:scatterings} we tested the origin of the photons that form the CS.
It has been shown that the CS, for a torus observed edge on, is mostly formed by \feka{} photons that have undergone the first four scatterings.
For column density $N_{\rm H}=10^{24.0}\,{\rm cm^{-2}}$ the different scattering events produced mostly symmetrical structures, and we did not observe a significant difference in their shapes.
Yet, for high column density $N_{\rm H}=10^{25.0}\,{\rm cm^{-2}}$, the first two scatterings are more asymmetrical than the higher order ones. We explored this further in Fig.\,\ref{fig:images_panels_scatterings}, where we show that the high asymmetry originates from the inner surface of the torus.
\item Dust: The last physical property tested, in \S\ref{sec:dust}, was the presence of dust grains in the torus.
The torus extends beyond the sublimation radius and thus, it is expected to be dusty.
We tested if the presence of iron on dust grains can change the shape of the CS. To do so, three different \textit{depletion factors} were used; 0, 0.7 and 1.0.
We found no significant differences neither to the shape of the CS nor in the fluxes and ratios (see\,Fig.\,\ref{fig:torus_dust}).
\item Velocity broadening: Given that many AGN can feature \feka{} emission line that have been broadened due to GR effects or the motion of the clouds where fluorescence happens, we have applied a gaussian smoothening at 6.4\,keV to see whether the CS is diluted by the broadening.
Although the kinematics of the circumnuclear material can be very complicated, we have decided to adopt a simple ad-hoc model to show that when any kind of disturbance is applied to the spectrum the CS is highly affected.
In \S\ref{sec:velbroad} are presented the results for four different velocities ($0-5000\, {\rm km/s}$) in two cases of \nh{} of the torus (see Fig\,\ref{fig:torus_velocity}).
It is clear that high broadening affect substantially the shape of the CS and therefore there are limitations to the use of the CS as a tool to study the material around the SMBH.
\item Geometries of the circumnuclear material: In \S\ref{geometries} different geometries were used to test the effect of different geometrical components on the shape and the flux of the CS.
We created six different models (M1-6) varying from a simple slab reflector to more complex geometries featuring an accretion disk, BLR, torus and a perpendicular hollow cone. Two different observing angles were considered. The edge on view ($90^{\circ}$) and mid $45^{\circ}$.
We found that the presence of the torus affects the shape of the CS (Models: M3-6), whereas the addition of the BLR (Models: M5-6) adds an extra layer of absorption and therefore affects the flux of the \feka{} and the CS. In the case of the simple slab reflector the flux is found to be lower with no significant contribution, whereas the spherical medium due to its isotropical nature it produces the same results in both observing angles as expected.
All the spectra for the different models are illustrated in\,Fig.\,\ref{fig:geometries}.
\end{itemize}

In \S\ref{sec:simulations} we used the complex model (M6), which includes the accretion disk, the BLR, the dusty torus and the polar medium (see\,Fig.\,\ref{fig:cartoon}), along with \textit{XRISM} and \textit{Athena} response files to generate simulated spectra for the Circinus galaxy. Velocity broadening of ${\rm FWHM}=1346\,{\rm km\,s^{-1}}$  \citep{Andonie+:2022} has been applied to the generated spectra using the \texttt{gsmooth} model of XSPEC.
Having the simulated spectra for a source that is known to have a \feka{} Compton shoulder, we were motivated to see whether we can retrieve the input parameters by fitting the CS using a grid model. 
The M6 model from \S\ref{geometries} is used. The properties of all the components except of the torus are fixed to the values of Table\,\ref{tab:component_details}. For the torus, the grid model has free parameters the column density and metallicity. Moreover, the velocity broadening and the spectral slope of the continuum are free to vary too.

Our test is focused on the parameters of the torus and our analysis shows that the model tends to slightly underestimate the FWHM, although the overall fit is good since the model used to generate the simulated spectra and the model used to fit them is the same. (see \S\ref{sec:simulation_properties}).
This is reported for both instruments (see Fig.\,\ref{fig:sim_xrism_properties} and Fig.\,\ref{fig:sim_athena_properties}).
The motivation was to illustrate the performance of advanced instruments in terms of spectral resolution and seek potential strong degeneracies between the properties.

Next, in \S\ref{sec:simulation_geom} we performed another test, this time focused on the geometrical configuration of the system, and whether there are degeneracies between different geometries.
We have applied three different models based on three different configurations.
The first model is a spherical cloud that surrounds the central source, which performs worse than the other two and suggests that it can be distinguished from the axisymmetric models.
When a simple torus along with the M6 model (see \S\ref{geometries}) are tested the fits gets better.
The two models are almost indistinguishable in terms of fit quality. 
When \textit{XRISM} data are tested the retrieved properties are the same except of the column density which is slightly larger for the simple torus model since it does not include the BLR which contribute to the line-of-sight \nh{}.
In the case of the \textit{Athena} spectrum the difference between the fit quality within the two models is larger, although no safe results can be derived from this analysis since the simulated spectra and the grid of the M6 model are identical.

We have shown that the Compton shoulder is a spectral feature sensitive to changes of the circumnuclear material.
Its shape is affected by some of the physical properties or the geometry of the reprocessing material.
Properties like the metallicity and the column density can have similar effects on the CS, however, they do not affect in a similar way both the \feka{} line and the CS.
With the new {\it XRISM}/Resolve instrument, it will be possible to use the \feka{} CS, along with broadband X-ray spectroscopy, to better constrain the physical properties and geometry of the circumnuclear material around AGN.

\section*{Acknowledgements}
The authors acknowledge the referees for they useful comments that improved this work.
GD acknowledges support from ANID Beca Doctorado Nacional 21211606.
CR acknowledges support from the Fondecyt Regular grant 1230345 and ANID BASAL project FB210003. 

\section*{Data availability}
All the data presented in this work were generated using the \Reflex{} code \citep{Paltani&Ricci:2017}. The data generated and/or analysed in this study are available from the corresponding author on reasonable request.
 
\bibliographystyle{mnras}
 \bibliography{compton_shoulder_mnras}

\appendix

\section{Different X-ray sources}\label{App:xray_source}

Throughout our analysis we used an X-ray input source modeled by a spherical corona, like the lamp-post model, with fixed a fixed power-law with a high energy cutoff (see, \S\ref{sec:xsource} for details).
However, the geometry of the X-ray source is still under debate, while AGN show a variety of spectral indices as well as energy cutoff \citep{Ricci+:2017Catalog}.
Hence, we performed some tests where the shape of the source deviates from the typical lamp-post to see whether it can affect the \feka{} line and its CS.
Moreover, we decided to explore two different spectral indices and check if it is likely to find differences that can be distinguishable.

\subsection{X-ray source geometry}\label{App:xray_source_geom}

A spherical corona that resides ontop of the SMBH has been used so far, with its properties illustrated in Table\,\ref{tab:source_details}.
Here we perform a series of simulations featuring different source geometries. For the reflecting medium a simple toroidal reflector similar to the one used in \S\ref{sec:physical_prop} has been selected, observed edge on.
First, we implement a simple point-like source that resides $10\,r_{\rm{g}}$ above the black hole.
It emits light in all directions.
Next, we change the geometry of the source by adding a sphere that is in the plane of the accretion disk that models an advection-dominated accretion flow (ADAF; \citealp{Yuan&Narayan:2014}), which can be found in low luminosity AGN.
An ADAF-like central source consists of very hot material that starts from radii close to ISCO (Innermost Stable Circular Orbit) and extends up to hundreds of $r_{\rm{g}}$. For our simulations we set center of the sphere in the center of the system and give it $100\,r_{\rm{g}}$ radius, which corresponds to $\rm \sim 1.62 \times 10^{-5}\,pc$.

\begin{figure}
    \centering
    \includegraphics[width=\columnwidth]{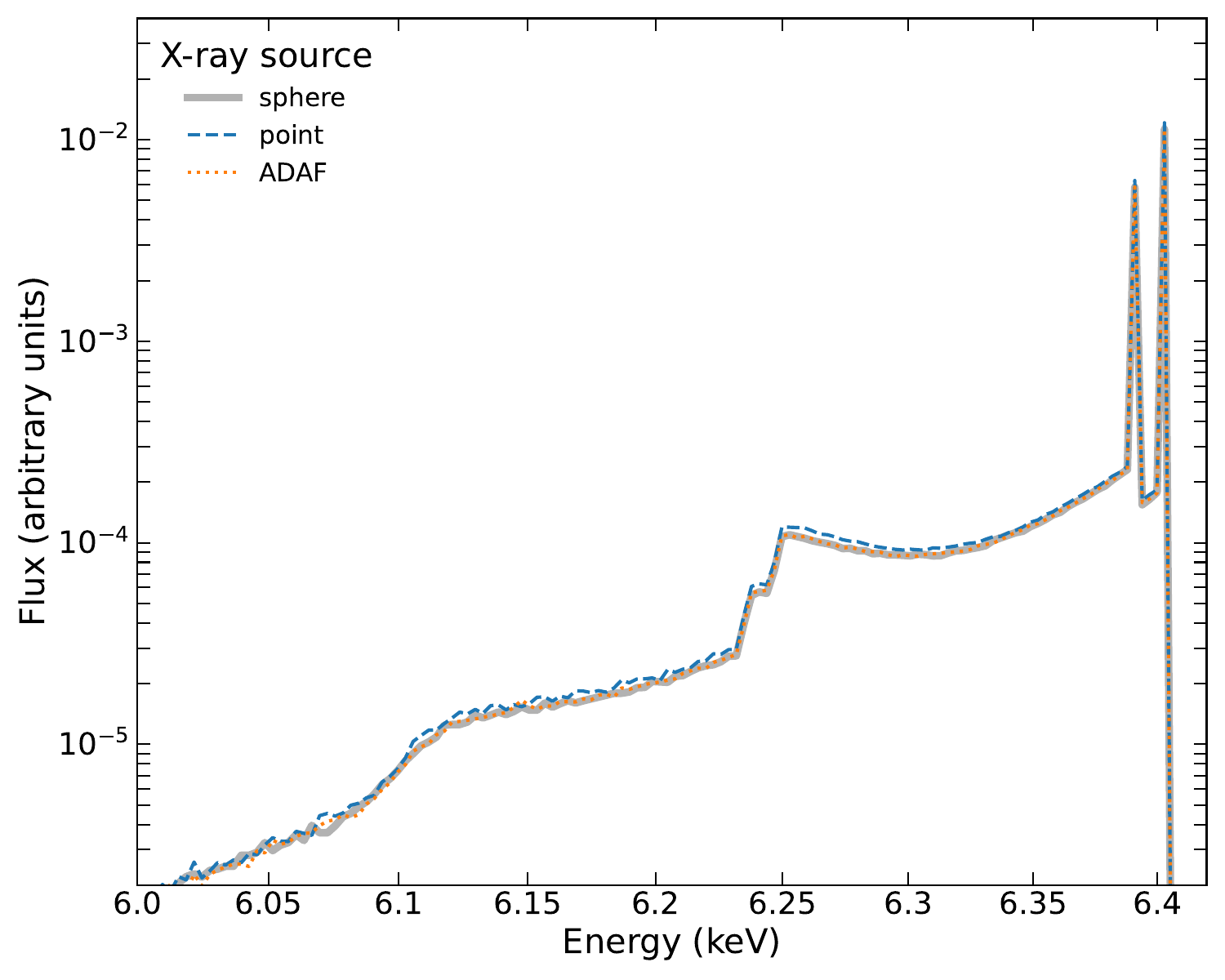}
    \vspace{-3mm}
    \caption{The \feka{} CS region for different type of X-ray source. Changing the geometry of the input source does not have any effect on the \feka{} and its CS as long as the spectral properties remain the same.}
    \label{fig:torusXsource}
\end{figure}

In\,Fig.\,\ref{fig:torusXsource} the Compton shoulder and \feka{} line are presented, for the point, the ADAF as well as the spherical corona source.
It is clear that the geometry and the position of source do not affect the shape of the CS.
We tested the CS ratio for the three different source and the difference is negligible ($<1\%$).
As a result, we suggest that the CS shape is independent of the geometry of the central X-ray source.

\subsection{X-ray source spectral index}\label{App:XsourceGamma}

\begin{figure}
    \centering
    \includegraphics[width=\columnwidth]{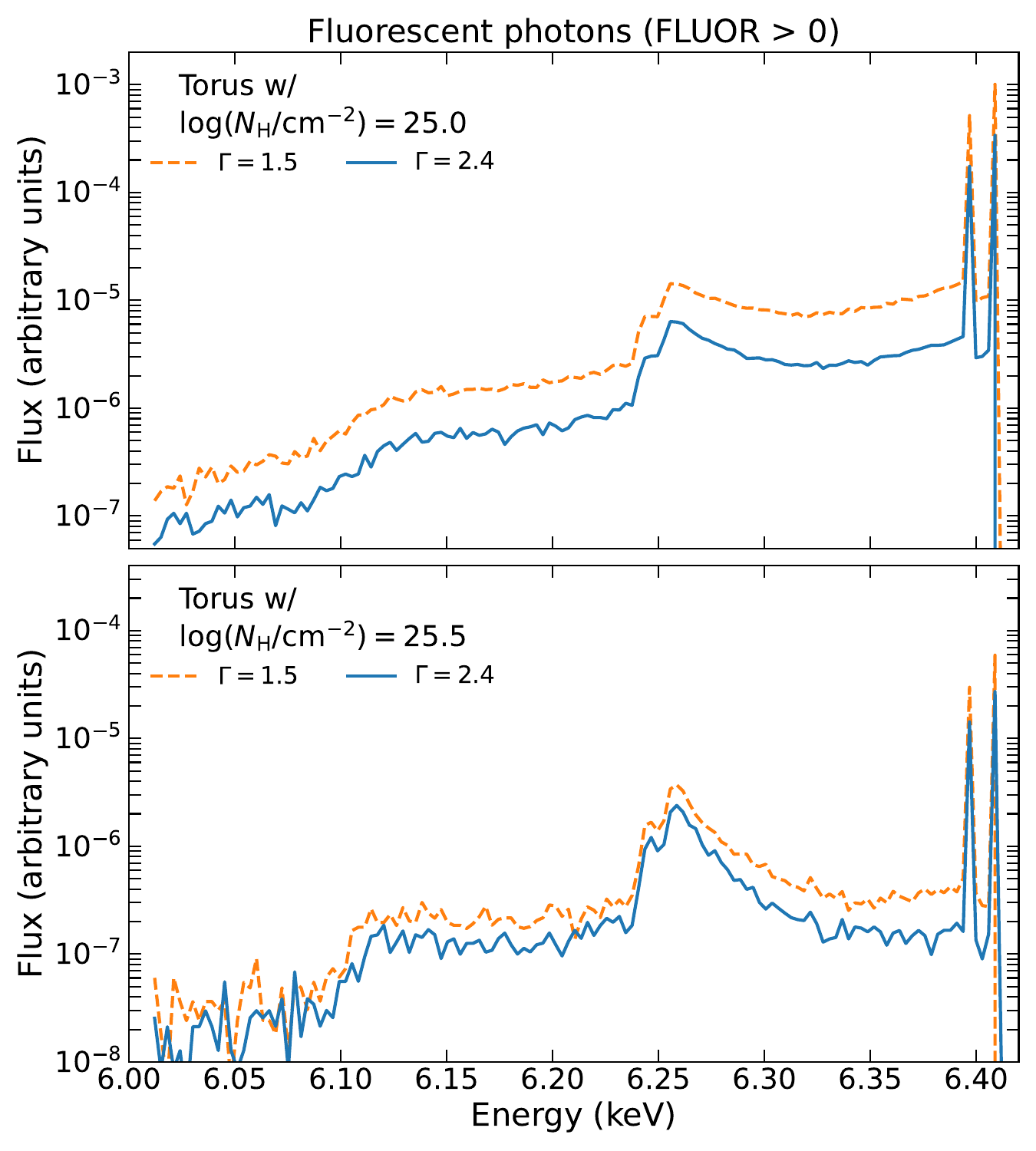}
    \vspace{-3mm}
    \caption{The \feka{} CS region for two different intercepting tori \textit{Top}: Column density $N_{\rm H}=10^{25.0}\,{\rm cm^{-2}}$, \textit{Bottom}: $N_{\rm H}=10^{25.5}\,{\rm cm^{-2}}$.Two opposite scenarios of spectral indices of the X-ray source have been tested. One flatter spectrum (1.5) and one steeper (2.4), which for a given distribution can affect the net amount of photons available near the K-edge. Only photons that originate in fluorescence are considered.
    }
    \label{fig:torus_gamma}
\end{figure}

In the simulations we have performed so far, the spectral index of the incident photon field is fixed to $\Gamma = 1.8$, which is common value for AGN \citep{Ricci+:2017Catalog}.
Nevertheless for a given power-law photon distribution any change on the spectral index affects the net amount of photons across the energy space.
Since the \feka{} emission line and the corresponding CS span in a very small energy range, we are interested in testing whether a different $\Gamma$ can produce different shapes of CS.
\citealt{Yaqoob&Murphy:2010} have performed a series of simulations to explore the connection between the spectral index and the shape of the CS.
They used a torus with column density $N_{\rm H}=1\times10^{25}\,{\rm cm^{-2}}$ and tested two different $\Gamma$ (1.5 - 2.5) for two observing angles; face-on and edge-on. First, in the face-on regime it is illustrated that the steepness of the of the continuum has a negligible effect on the shape of the CS which in fact makes it non-detectable.
However, in the edge-on case the two spectral indices produce distinct spectral shapes of the CS. Flatter spectra provide a larger amount of high energy photons that can penetrate deeper to the torus and therefore are more likely to produce a more symmetrical CS as a result of wider range of scattering angles. On the other hand, steeper spectra ($\Gamma = 2.5$) show a significant back-scattering peak due to the fact that the CS emerges from the back surface of the torus with lesser photons being able to penetrate deep into the medium. 
\citealt{Odaka:2016} used a spherical cloud as circumnuclear medium, featuring column density $N_{\rm H}=1\times10^{24}\,{\rm cm^{-2}}$ and metallicity $Z=1$, to test a variety of spectral indices ($1.2-2.8$).
For the tested column density they did not find any particular difference in the shape of the \feka{} and the CS. They report that the equivalent width of both features ($EW_{\rm FeK\alpha},\ EW_{\rm CS}$) show an anti-correlation with the photon index, which suggests that $\Gamma$ affects the production of fluorescent photons and not the shape of the CS, which is the outcome of scatterings.
However, the column density they tested it might not be enough for the CS to show a difference in its shape for steeper and flatter continua.
We decided to apply a similar exercise by testing two opposite scenarios of $\Gamma$ in two tori with high column densities ($\log (N_{\rm{H}}/{\rm cm^{-2}}) = 25.0 - 25.5$). The spectra are collected for edge-on observing angles, since it has already been shown that for face-on angles the effect is negligible. The properties of the torus are fixed to the properties presented in \S\ref{sec:torus}. In Fig.\,\ref{fig:torus_gamma} two panels are presented for the two different column densities. In both cases are illustrated the spectra in the \feka{} energy region for a soft continuum ($\Gamma = 2.4$, solid line) and a hard one ($\Gamma = 1.5$, dashed line).
For $\Gamma = 1.5$ in both panels it is clear that the total flux of the \feka{} and the CS is higher than $\Gamma = 2.4$ due to the larger amount of photons that can cause fluorescence. 
Considering the shape of the CS, we do not observe any significant variations in none of the two cases tested, suggesting that the formation of more or less \feka{} photons dominates any scattering process. Therefore, the Compton shoulder cannot be a reliable feature to constrain neither the geometry nor the spectral index of the X-ray source since they do not have a strong effect on it.

\section{The Broad Line Region}\label{App:BLRblocks}

\begin{figure}
    \centering
    \includegraphics[width=\columnwidth]{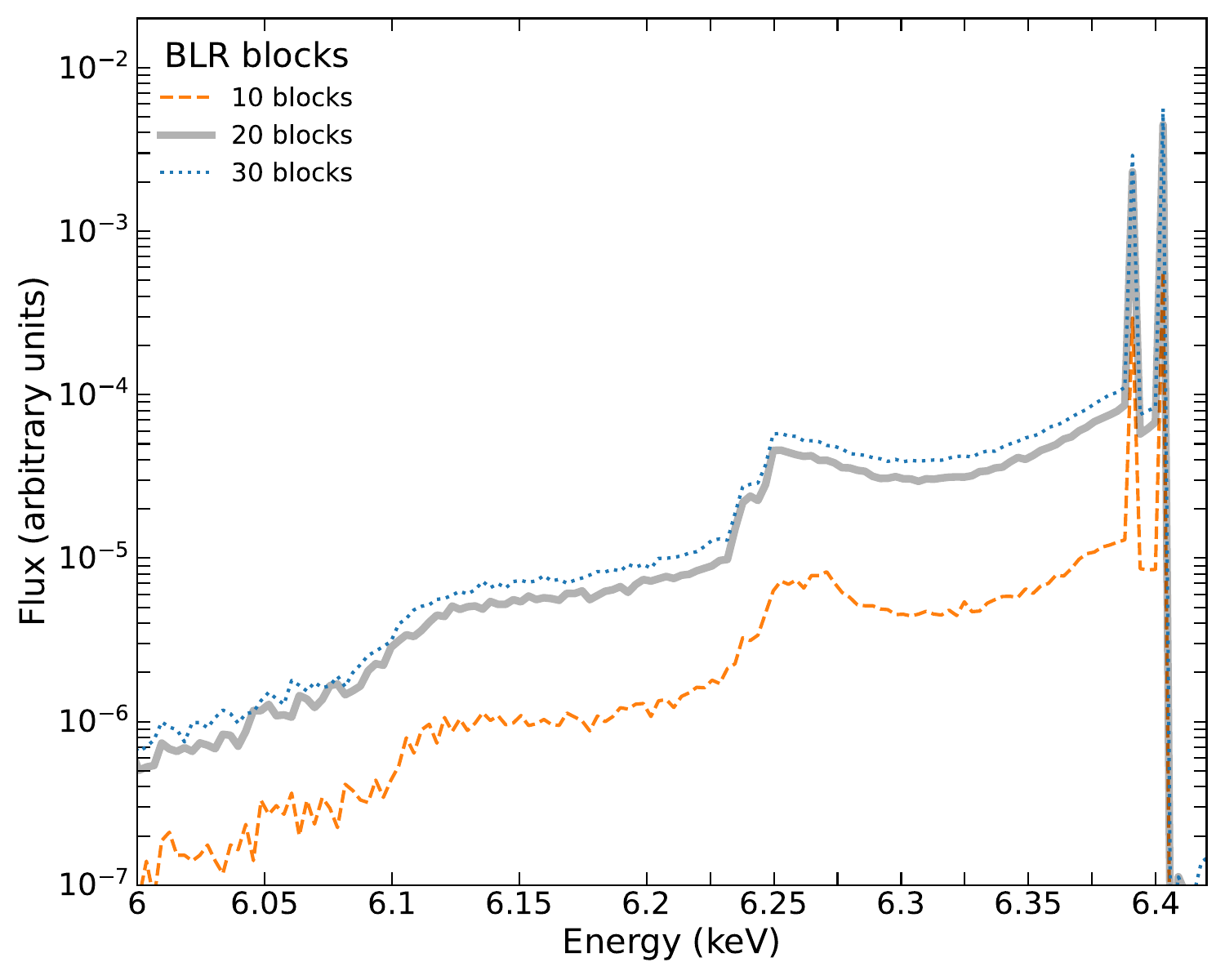}
    \vspace{-3mm}
    \caption{The CS spectral region for BLR consisting of different number of building blocks. Higher complexity increases the computational time of our simulations and therefore we select 20 blocks that provide sufficient photons while keeping the simulations in reasonable running times.}
    \label{fig:number_blocks}
\end{figure}

In\,Fig.\,\ref{fig:cartoon} a schematic of all the components, used, is presented.
The broad line region (BLR) has been modeled as a flared disk; a disk with increasing z-axis height as we move across its radius.
In order to build the flared disk we used a combination of blocks one on top of each other, with decreasing radial length. To test the efficiency of this method we test BLR models that consist of 10, 20 and 30 blocks. The cubic density was fixed in all blocks (see\,Table\,\ref{tab:component_details}).
In\,Fig.\,\ref{fig:number_blocks} the spectra for different number of blocks are illustrated.
In case of 10 blocks the total flux of the \feka{} and its CS is low. On the other hand, when the BLR consists of 20 and 30 blocks the difference is negligible.
Consequently, we decided to use 20 blocks in the rest of our work, a number that saves computational time and provides good resolution.

\section{\h2{} Fraction}\label{App:h2fraction}
\begin{figure}
    \centering
    \includegraphics[width=\columnwidth]{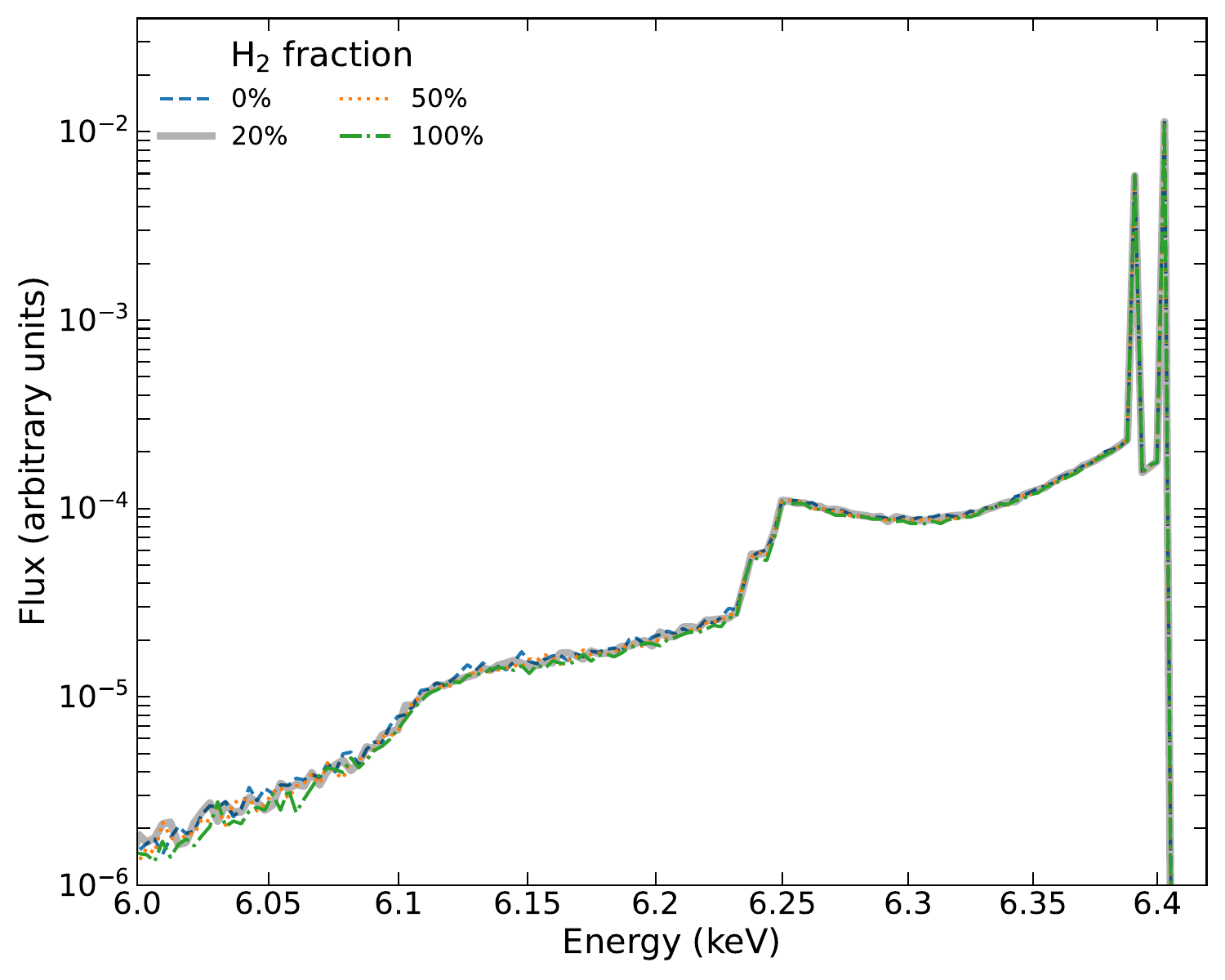}
    \vspace{-3mm}
    \caption{The \feka{} CS spectral region for different H2 fractions. The amount of molecular hydrogen in the circumnuclear material has no effect on the CS and the \feka{} line.}
    \label{fig:torus_H2}
\end{figure}

Another physical parameter tested is the molecular hydrogen percentage in the circumnuclear material (\h2{} fraction). By tuning \h2{} option in our simulations we can change the percentage of \h2{} that is present in the reprocessing medium. We set four different simulations, with increasing $H_2$; i) 0, ii) 0.2, iii) 0.5, iv) 1.0 using a torus with the default settings as presented in Table\,\ref{tab:component_details}. In\,Fig.\,\ref{fig:torus_H2} the spectra of the simulations are presented. We can see that the \h2{} percentage does not affect the shape of the CS, whatsoever. \citet{Sunyaev:1996} have shown that molecular hydrogen and helium interact mostly elastically with the X-ray photons and therefore they do not affect the Compton shoulder.

\section{Inclination angle}\label{App:Inclination}

The main analysis of this work is focused on the circumnuclear material and how its physical properties or geometrical configuration can affect the \feka{} and its corresponding Compton shoulder.
Hence, the observing angle, a property that it is an effect that it is related to the relative orientation between the observer and the source has not been explored in detail (except of Fig.\,\ref{fig:geometries}).
Nevertheless it is known that that various observing angles can indeed produce different CS shapes.
Since not all AGN are observed edge on, we test how the \feka{} and CS vary with the observing angle.
\citet{Odaka:2016} used a slab geometry reflector with an X-ray source above it. They collect the spectrum from 5 different angle bins ($\rm 0^\circ-10^\circ,\ 20^\circ-30^\circ,\ 40^\circ-50^\circ,\ 60^\circ-70^\circ,\ 80^\circ-90^\circ$).
In angles $\rm \geq 20^\circ$ the shape of the shoulder is relatively the same. In the face-on regime there is a peak at $\rm \sim 6.32\, keV$, which corresponds to a scattering angle of $\rm 90^\circ$.

\begin{figure}
    \centering
    \includegraphics[width=\columnwidth]{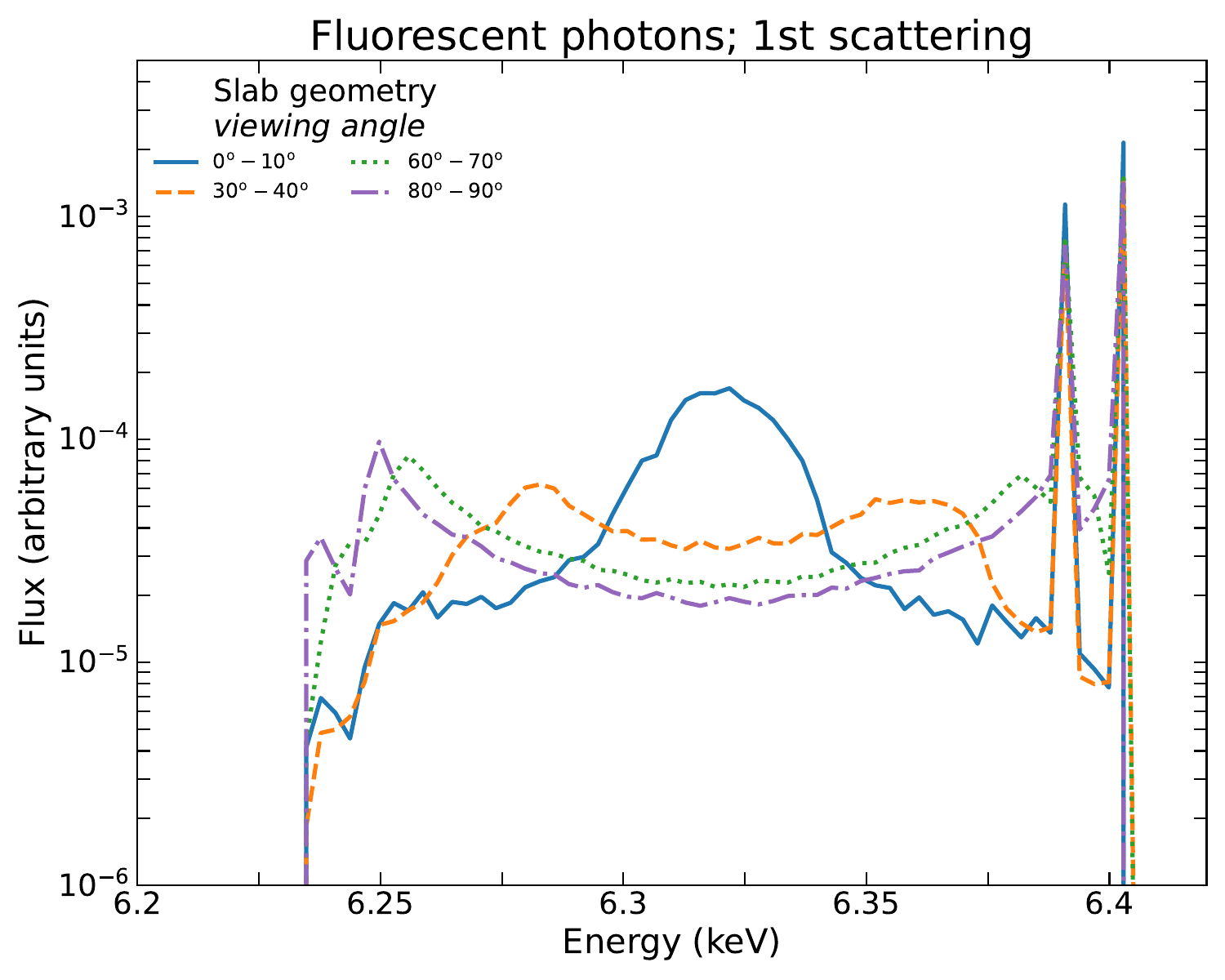}
    \vspace{-3mm}
    \caption{\textit{Top}: Spectra in the range of the \feka{} line and its CS for a slab reflector observed in different angles. The spectra illustrate photons that originate from fluorescence and have undergone one scattering.}
    \label{fig:slab_angle}
\end{figure}

In our simulations, we build a simple model that includes an X-ray source (see\,Table\,\ref{tab:source_details} and a slab reflector (disk in Table\,\ref{tab:component_details}).
To focus only on the \feka{} line and its corresponding CS, we collect photons that come from fluorescence and have been scattered only once ($\rm SCATT == 1$).
We have performed 4 different simulations collecting photons in different angles covering from face-on to edge-on scenario.
In\,Fig.\,\ref{fig:slab_angle} the \feka{} and CS region is shown. Our results are in agreement with \citet{Odaka:2016} and illustrate the dominant scattering angle in each case and, consequently, the amount of energy a fluorescent photon loses after its first Compton scattering.

\begin{figure}
    \centering
    \begin{subfigure}
        {\columnwidth}\includegraphics[width=\columnwidth]{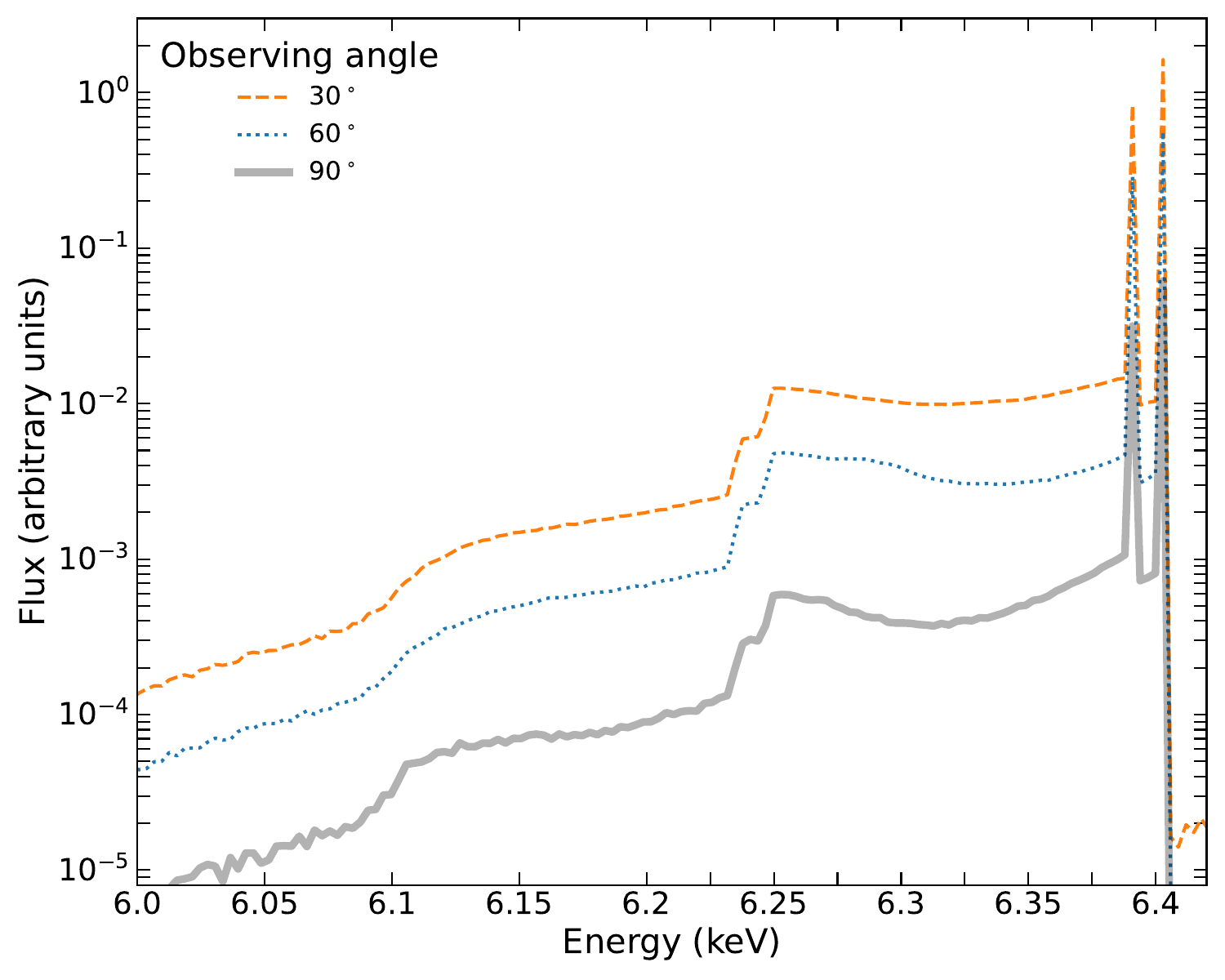}
    \end{subfigure}
    \hfill
    \begin{subfigure}
        {\columnwidth}\includegraphics[width=\columnwidth]{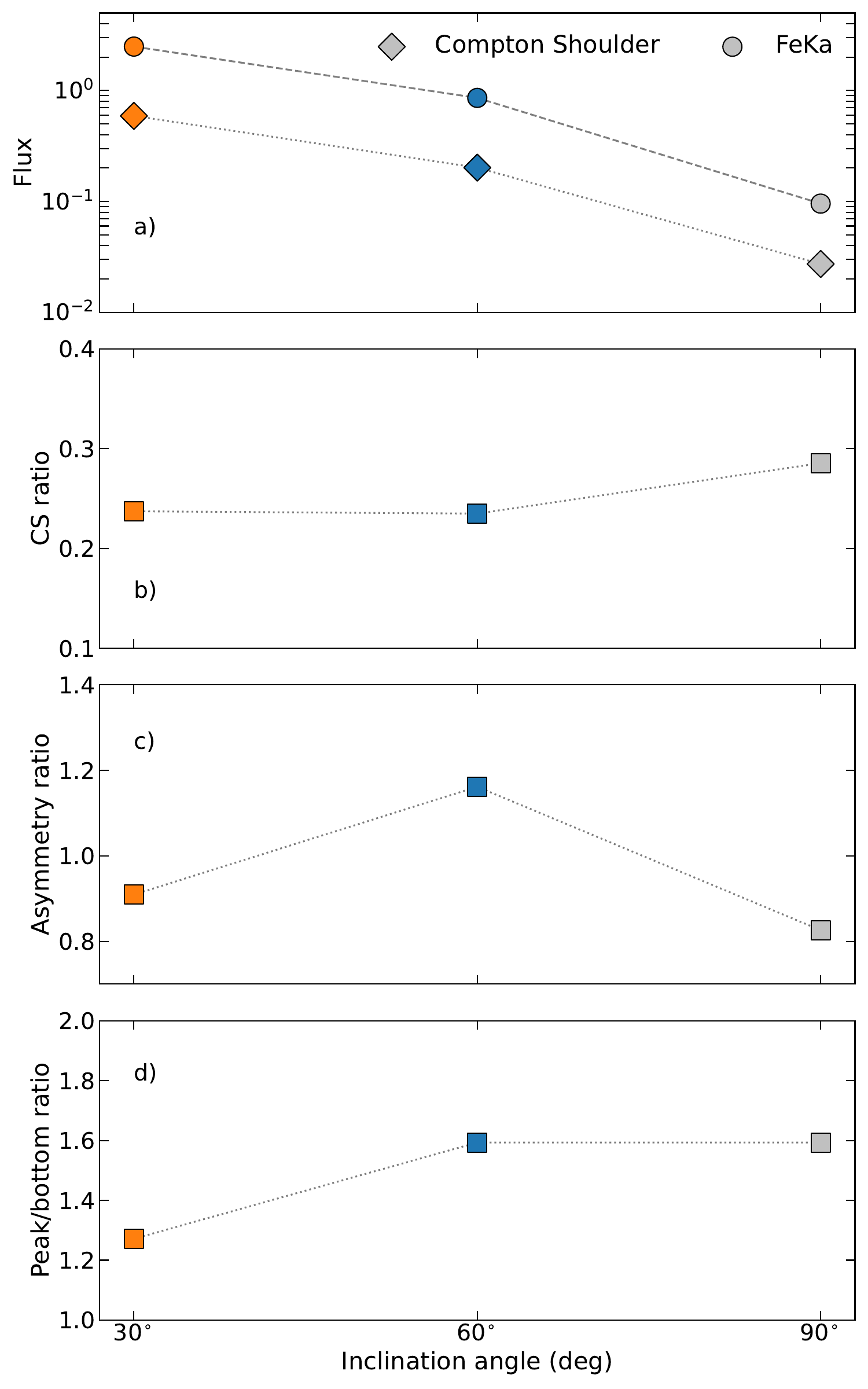}
    \end{subfigure}
    \vspace{-3mm}
    \caption{\textit{Top}: The AGN model (M6) observed in different inclination angles. \textit{Bottom}: a) The CS and \feka{} flux, b) The ratio of the CS flux and the \feka{} line, c) The asymmetry ratio and d) peak/bottom ratio. The colors represent the column densities introduced in the top panel.}
    \label{fig:AGN_angle}
\end{figure}

Having tested how the shape of the CS changes in the case of a slab reflector, we proceed to test different observing angles for more complex geometries. To do so, we use the M6 model presented in \S\ref{geometries} which includes all of the components from Table\,\ref{tab:component_details}. We set three different inclination angles for the simulations; $\rm 30^\circ,\ 60^\circ,\ 90^\circ$ ($\rm 5^{\circ}\, bin$).
In\,Fig.\,\ref{fig:AGN_angle} the spectra are presented in the \feka{} energy region.

In panel (a) of Fig.\,\ref{fig:AGN_angle} it is shown that the total flux drops with the angle as a result of the absorption.
Moreover, the shape of the CS changes, too.
Low inclination angles produce a much flatter CS, which can be related to the previous simulation where the disk in low angles enhance the shoulder. The shoulder becomes asymmetrical for $\rm 60^\circ$ and then returns to a more symmetrical profile (see panel (c) of Fig.\,\ref{fig:AGN_angle}). At the same time, the peak at $\rm 6.24\, keV$ becomes more apparent as the angle increases.
In\,Fig \ref{fig:AGN_angle} panel (d) the peak/bottom ratio shows a positive trend.

\label{lastpage}

 \end{document}